\documentclass[12pt,3p,doublespacing]{elsarticle}
\journal{.}
\usepackage{times}
\usepackage{libertine}
\emergencystretch=\maxdimen
\usepackage{sectsty}
\sectionfont{\color{blue}\large\MakeUppercase}
\subsectionfont{\color{blue}}
\biboptions{sort&compress}
\usepackage[colorlinks]{hyperref}
\AtBeginDocument{%
	\hypersetup{
		plainpages = false, 
		bookmarksopen = true,
		bookmarksnumbered = true,
		breaklinks = true,
		linktocpage,
		colorlinks = true,
		linkcolor = purple,
		urlcolor  = black,
		citecolor = Magenta,
}}
\usepackage{multicol}
\usepackage{multirow}
\usepackage{imakeidx}
\makeindex
\usepackage{nomencl}
\makenomenclature
\usepackage[xindy]{glossaries}
\makeglossaries
\usepackage{graphics}
\usepackage{graphicx}
\usepackage{epsf,psfrag}
\usepackage{epstopdf}
\usepackage[para]{threeparttable}
\usepackage{subfigure}
\usepackage{epsfig}
\usepackage{pdflscape}
\usepackage{rotating}
\usepackage{lscape}
\usepackage{float}
\usepackage{stfloats}
\usepackage{textgreek}
\usepackage{longtable}
\usepackage{lscape}
\usepackage{capt-of}
\usepackage{tablefootnote}
\usepackage{footnote}
\usepackage{caption}
\usepackage[autopunct=true]{csquotes}
\usepackage{rotating}
\usepackage[normalem]{ulem}
\usepackage{resizegather}
\usepackage{siunitx}
\usepackage[nodisplayskipstretch]{setspace}
\usepackage{color,soul}
\usepackage[usenames,dvipsnames]{xcolor}
\usepackage{amsfonts,amsmath,amssymb,gensymb}
\usepackage{latexsym,array}
\usepackage{textcomp}

\newcommand{\RomanNumeralCaps}[1]
\linenumbers
\makesavenoteenv{tabular}
\makesavenoteenv{table}
\usepackage[left,displaymath, mathlines]{lineno}
\DeclareMathAlphabet{\mathpzc}{OT1}{pzc}{m}{it}
\def\fig{Fig.~}

\def\eqn{Eq.~}
\def\eqns{Eqs.~}
\def\tab{Table~}

\newcommand{\myvec}[1]{\mathbf{#1}}     
\def\tsc#1{\csdef{#1}{\textsc{\lowercase{#1}}\xspace}}
\tsc{WGM}
\tsc{QE}
\tsc{EP}
\tsc{PMS}
\tsc{BEC}
\tsc{DE}
\usepackage{easyReview} \setreviewsoff
\newcommand{\removeEq}[1]{\ifistoreview{\@\expandafter\removeColor{#1}\hspace{-0.6em}} \else {}\fi}
\newcommand{\ttxt}[1]{\textbf{\color{blue}\large\MakeUppercase{#1}}}     

\graphicspath{{../}{Figures/}{Figures/Validation/}}
\DeclareGraphicsExtensions{.png, .PNG,.pdf,.PDF,.jpg,.JPG,.eps,.EPS}
\begin{document}

%
\setcounter{page}{1}
\begin{frontmatter} 
%
%
%
%
%
\title{\ttxt{CFD analysis of electroviscous effects in electrolyte liquid flow through heterogeneously charged non-uniform microfluidic device}}
\author[labela]{Jitendra {Dhakar}}
\author[labela]{Ram Prakash {Bharti}\corref{coradd}}\ead{rpbharti@iitr.ac.in}
\address[labela]{Complex Fluid Dynamics and Microfluidics (CFDM) Lab, Department of Chemical Engineering, Indian Institute of Technology Roorkee, Roorkee - 247667, Uttarakhand, INDIA}
%
%
\cortext[coradd]{\textit{Corresponding author. }}
%
\begin{abstract}
\fontsize{11}{16pt}\selectfont
In electrokinetic flows, charge-heterogeneity (i.e., axial variation of surface charge in the microfluidic device) affects the microfluidic hydrodynamics for practical applications such as mixing, heat, and mass transfer processes. In this work, pressure-driven flow of symmetric (1:1) electrolyte liquid through heterogeneously charged contraction-expansion (4:1:4) microfluidic device, accounting the influence of electroviscous effects has been investigated numerically. Total potential ($U$), ion concentrations ($n_\pm$), velocity ($\myvec{V}$), and pressure ($P$) fields are obtained after solving the mathematical model consisting of the Poisson's, Nernst-Planck (NP), Navier-Stokes (NS), and continuity equations numerically using the finite element method (FEM). Results are presented for wide ranges of dimensionless parameters such as inverse Debye length ($2\le K\le 20$), surface charge density ($4\le S_\text{1}\le 16$), and surface charge-heterogeneity ratio ($0\le S_\text{rh}\le 2$). Results show that the total potential ($\Delta U$) and pressure ($\Delta P$) drops change maximally by 3511.45\% \add{(0.2127 to 7.6801)} (at $S_1=4$, $K=20$) and 41.4\% \add{(1.0941 to 1.5471)} (at $S_1=16$, $K=2$), respectively with overall enhancing charge-heterogeneity ($0\le S_\text{rh}\le 2$), over the ranges of $K$ and $S_1$. Electroviscous correction factor, $Y$ (i.e., ratio of apparent to physical viscosity) increases maximally by 24.39\% \add{(1.1158 to 1.3879)} (at $K=4$, $S_\text{rh}=1.75$), 37.52\% \add{(1.0597 to 1.4573)} (at $S_1=16$, $S_\text{rh}=2$), and 41.4\% \add{(1.0306 to 1.4573)} (at $S_1=16$, $K=2$) with the variation of $S_1$ from 4 to 16, $K$ from 20 to 2, and $S_\text{rh}$ from 0 to 2, respectively. Further, overall increment in $Y$ is noted as 45.73\% \add{(1 to 1.4573)} (at $K=2$, $S_1=16$, $S_\text{rh}=2$), relative to non-EVF ($S_1=0$ or $K=\infty$). Thus, charge-heterogeneity enhances electroviscous effects in microfluidic devices, which enables the uses of present numerical results for designing the reliable and essential micro-sized channels for practical microfluidic applications.
\end{abstract}
\begin{keyword}
\fontsize{11}{13pt}\selectfont
Electroviscous effect\sep Pressure-driven flow\sep Charge-heterogeneity\sep Microfluidic device
\end{keyword}
\end{frontmatter}
%
\section{Introduction}
\label{sec:intro}
\noindent Microfluidic devices have gained the significant importance over the years for their broader ranges of practical applications such as micro heat pump, micro heat sink, drug delivery system, DNA analysis \citep{bhushan2007springer,li2008encyclopedia,lin2011microfluidics,nguyen2013design,bruijns2016microfluidic,li2021microfluidic}. Surface forces, i.e., electrical, surface tension, magnetic forces, etc. significantly affect the fluid dynamics in the micro-scale devices \citep{hunter2013zeta,li2001electro}, therefore, it depicts the different flow characteristics as compared to macro-scale flow. At micro-scale, understanding the electrokinetic phenomena is essentially important for developing the reliable and efficient microfluidic devices for practical applications such as mixing, heat, and mass transfer processes. 

\noindent The electroviscous (EV) flow in the electrokinetic phenomena develops due to applied pressure-driven flow (PDF) of electrolyte liquid in the charged microfluidic device. Charged surfaces of device attract counter-ions, thus, rearrangement of the ions close to the solid-liquid interface forms an `electrical double layer' (EDL) which consists of Stern and diffuse layers (refer \fig\ref{fig:1}) \citep{li2001electro,davidson2007electroviscous,dhakar2022electroviscous,dhakar2023cfd}. The flow of mobile ions in diffuse layer due to imposed PDF on the liquid, generates a convection current which is known as `streaming current' ($I_\text{s}$) (refer \fig\ref{fig:1}). The potential related to this current is called as streaming potential, which imposes a `conduction current` ($I_\text{c}$) in the opposite direction of the PDF (refer \fig\ref{fig:1}). It drives liquid with them in the opposite direction of the primary flow and reduces the net flow of electrolyte liquid. This phenomena is commonly known as the `electroviscous effect' \citep{hunter2013zeta,atten1982electroviscous,dhakar2022electroviscous,dhakar2023cfd}.

\noindent Over the years, electroviscous effects have studied broadly in the microfluidic devices and detailed literature is summarized by \citet{dhakar2022electroviscous,dhakar2023cfd}. Experimental and numerical investigations are carried out to explore the electroviscous effects in pressure-driven liquid flow through symmetrically charged ($S_\text{r}=1$) microfluidic devices with uniform cross-sections such as cylinder \citep{rice1965electrokinetic,levine1975theory,bowen1995electroviscous,brutin2005modeling,bharti2009electroviscous,jing2016electroviscous}, slit \citep{burgreen1964electrokinetic,mala1997flow,mala1997heat,chun2003electrokinetic,ren2004electroviscous,chen2004developing,joly2006liquid,xuan2008streaming,wang2010flow,jamaati2010pressure,zhao2011competition,tan2014combined,jing2015electroviscous,matin2016electrokinetic,jing2017non,matin2017electroviscous,kim2018analysis,sailaja2019electroviscous,mo2019electroviscous,li2021combined,li2022electroviscous,banerjee2022analysis}, rectangular \citep{yang1998modeling,li2001electro,ren2001electro}, and elliptical \citep{hsu2002electrokinetic} as well as non-uniform cross-sections such as contraction-expansion cylinder \citep{bharti2008steady,davidson2010electroviscous}, slit \citep{davidson2007electroviscous,berry2011effect}, and rectangular \citep{davidson2008electroviscous}. The influence of contraction ratio and electroviscous effects on the electrolyte liquid flow through symmetrically charged microchannel are analyzed recently \citep{dhakar2023the}. However, \citet{dhakar2023cfd} have analyzed electroviscous flow in the asymmetrically charged ($S_\text{r}\neq1$) contraction-expansion microfluidic device. These studies \citep{davidson2007electroviscous,davidson2008electroviscous,bharti2008steady,bharti2009electroviscous,davidson2010electroviscous,dhakar2022slip,dhakar2022electroviscous,dhakar2023cfd,dhakar2023the} concluded that the dimensionless parameters such as surface charge density ($4\le S\le 16$), inverse Debye length ($2\le K\le 20$), surface charge ratio ($0\le S_\text{r}\le 2$), contraction ratio ($0.25\le d_\text{c}\le 1$) and slip length ($0\le B_\text{0}\le 0.20$) strongly influence the flow characteristics, i.e., total potential ($U$), induced electric field strength ($E_\text{x}$), excess charge ($n^\ast$), and pressure ($P$) fields in the microfluidic devices. Further, simple pseudo-analytical model \citep{davidson2007electroviscous,davidson2008electroviscous,bharti2008steady,bharti2009electroviscous,davidson2010electroviscous,dhakar2022slip,dhakar2022electroviscous,dhakar2023cfd,dhakar2023the} has been developed to predict the pressure drop and electroviscous correction factor in the symmetrically/asymmetrically charged contraction-expansion microfluidic devices, which overestimated the pressure drop maximally by 5--10\% compared to their numerical results.

\noindent Surface heterogeneity is significant characteristic of the microfluidic device which can occur because of chemical species absorption by surface \citep{ghosal2003effect}, surface treatment defects \citep{ajdari1996generation}, and controlling surface charge distribution \citep{jain2013optimal,bhattacharyya2019enhanced}. Surface charge heterogeneity affects the practical microfluidic applications like mixing efficiency \citep{nayak2018mixing,chu2019magnetohydrodynamic,guan2021mixing}, mass and heat transfer rates \citep{ghosal2006electrokinetic,ng2012dispersion,azari2020electroosmotic} in microfluidic devices. `Charge-heterogeneity' (CH) is defined as the axial variation of surface charge in the microfluidic devices, i.e., two or more surfaces constructed by different materials  are connected in the series manner.

\noindent Further, fewer attempts are made to investigate electroviscous flow through heterogeneously charged microfluidic devices. \citet{xuan2008streaming} has analytically studied electroviscous effect and streaming potential in uniform microchannel, accounting surface charge variations such as parallel ($q\parallel\nabla P$) and perpendicular ($q\perp\nabla P$) to the external applied pressure gradient. He concluded that the electroviscous effect and streaming potential are dependent on the arrangement of surface charge heterogeneity for smaller $K<50$ and such dependence becomes weak for larger $K>50$ \citep{xuan2008streaming}. Recently, \citet{dhakar2023analysis} have investigated the electroviscous flow of electrolyte liquid through heterogenously charged uniform slit microchannel. They have shown that the charge-heterogeneity ($0\le S_\text{rh}\le 2$) strongly influences the flow fields (i.e., total potential, excess charge, induced electric field strength, and pressure fields) in the uniform microfluidic device. 

\noindent As the best of our knowledge, charge-heterogeneity effects in the electroviscous flow are unexplored in the literature in non-uniform geometries. This study has analyzed the electroviscous effects in pressure-driven flow of electrolyte liquid through heterogeneously charged contraction-expansion microfluidic device. A finite element method (FEM) is used to solve the governing equations such as the Poisson's, Nernst-Planck, and Navier-Stokes equations to obtain flow fields, i.e., total potential ($U$), ion concentrations ($n_\pm$), velocity ($\myvec{V}$), and pressure ($P$) fields in the microchannel for given ranges of dimensionless parameters ($2\le K\le 20$, $4\le S_\text{1}\le 16$, $0\le S_\text{rh}\le 2$).
\section{Mathematical formulation}
\subsection{Problem description}
\begin{figure}[h]
	\centering
	\includegraphics[width=1\linewidth]{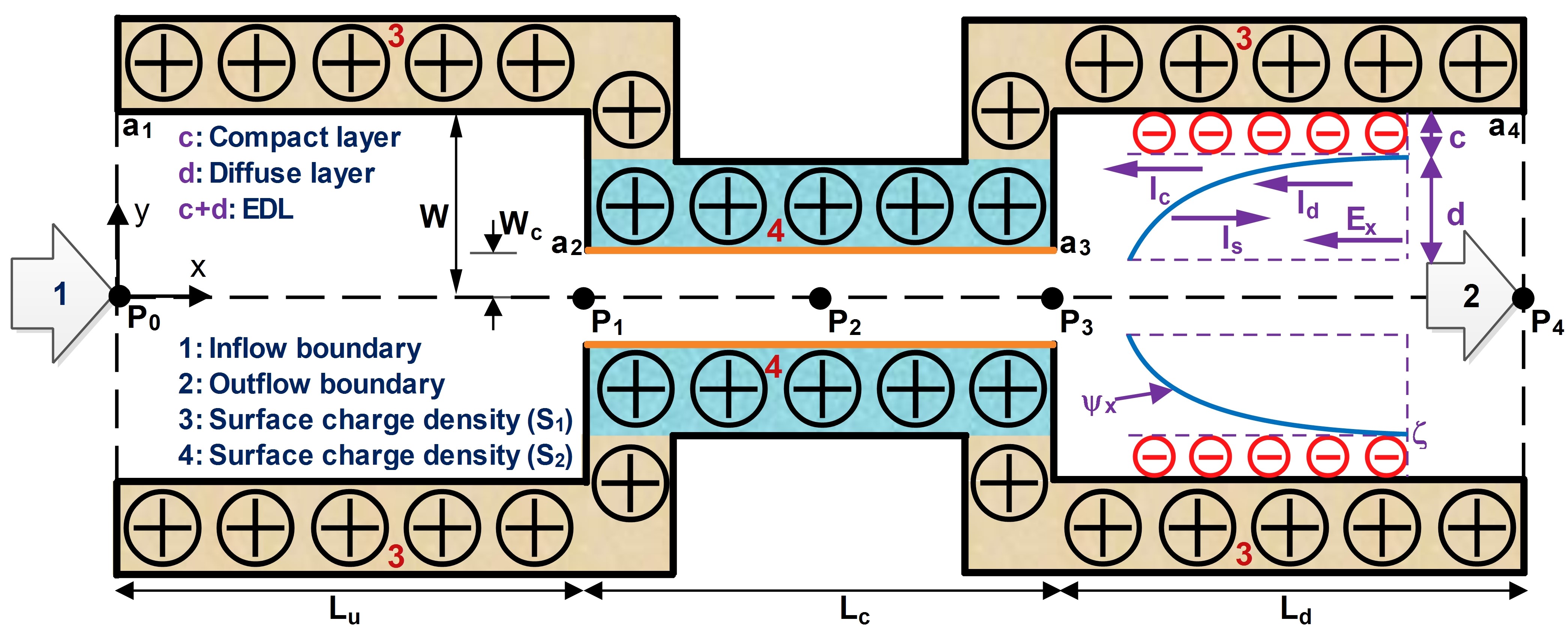}
	\caption{Illustration of electroviscous flow (EVF) through heterogeneously charged non-uniform microfluidic geometry.}
	\label{fig:1}
\end{figure}

\noindent \fig\ref{fig:1} illustrates electroviscous flow (EVF) of electrolyte liquid through heterogeneously charged contraction-expansion (4:1:4) two-dimensional (2-D) slit microfluidic device. Flow is assumed as laminar and fully developed (with an average flow velocity $\bar{V}$, m/s) in the microfluidic device. Contraction section is situated between the upstream and downstream sections. The sizes (in $\mu$m) of upstream, contraction, and downstream sections are ($L_\text{u}$, $2W_\text{u}$), ($L_\text{c}$, $2W_\text{c}$), and ($L_\text{d}$, $2W_\text{d}$), respectively. Total length (in $\mu$m) of device is expressed as $L=L_\text{u}+L_\text{c}+L_\text{d}$ and contraction ratio is defined as $d_\text{c}=W_\text{c}/W$. Uniform but unequal surface charge densities (in C/m$^2$) are assumed on the walls of upstream/downstream and contraction sections of the device as $\sigma_{\text{1}}$ and $\sigma_{\text{2}}$, respectively (refer \fig\ref{fig:1}). The geometric mean concentration of each ion species is $n_0$ \citep{harvie2012microfluidic,davidson2016numerical,dhakar2022electroviscous}. The electrolyte liquid is considered to have symmetric ($1$:$1$) cations and anions with equal ion valances ($z_{{+}}=-z_{{-}}=z$) and diffusivity ($\mathcal{D}_{+}=\mathcal{D}_{-}=\mathcal{D}$, m$^2$/s). The electrolyte liquid is assumed to be Newtonian and incompressible, i.e., viscosity ($\mu$, Pa.s), density ($\rho$, kg/m$^3$), and dielectric constant ($\varepsilon_{\text{r}}$) are spatially uniform. The dielectric constant of the device wall is assumed to be very small compared to liquid ($\varepsilon_{\text{r,w}} \lll \varepsilon_{\text{r}}$).
\subsection{Governing equations and boundary conditions}
\noindent The mathematical formulation of electroviscous (EV) flow mathematical model can be expressed by the conservation of potential, mass of each ion species, mass and momentum by the Poisson's, Nernst-Planck (NP), Navier-Stokes (NS) and continuity equations. The governing equations in the dimensional form are explained and presented elsewhere \citep[refer \eqns A.1 to A.16 in][]{dhakar2022electroviscous}. Scaling parameters such as $\bar{V}$, $\rho\bar{V}^2$, $n_{\text{0}}$, $U_\text{c}=(k_{\text{B}}T/ze)$, $W$, and $W/\overline{V}$ for velocity, pressure, number density of ions, electrical potential, length, and time, respectively are used to non-dimensionlization of these governing equations. The dimensionless groups obtained from the scaling analysis are given below.
\begin{gather}
	Re=\frac{\rho\bar{V}W}{\mu}, \qquad
	\mathit{Sc}=\frac{\mu}{\rho \mathcal{D}}, \qquad
	Pe =Re\times\mathit{Sc}, \qquad
	\beta=\frac{\rho\varepsilon_{\text{0}}\varepsilon_{\text{r}}U^2_\text{c}}{2\mu^2}, \qquad 
	K^2=\frac{2W^2zen_{\text{0}}}{\varepsilon_{\text{0}}\varepsilon_{\text{r}}U_\text{c}}
	\label{eq:1}
\end{gather}
where $Re$, $\mathit{Sc}$, $Pe$, $\beta$, and $K$ are the Reynolds, Schmidt, Peclet number, liquid parameter, and inverse Debye length, respectively. Further, $W$, $\rho$, $\bar{V}$, $\mu$, $\mathcal{D}$, $e$, $z$, $k_\text{B}$, $T$, $\varepsilon_{\text{0}}$, $\varepsilon_{\text{r}}$, and $n_0$ are the half-width of channel, density of liquid, average velocity, viscosity of liquid, diffusivity of ions, electron charge, valances of ions, Boltzmann constant, temperature, permittivity, dielectric constant of liquid, and geometric mean ion concentration, respectively.

\noindent The dimensionless governing equation and boundary conditions (i.e., \eqns\ref{eq:2} to \ref{eq:8}) for present physical problem are expressed as follows (variables names retain same in as dimensional equations \citep[refer \eqns A.1 to A.16 in][]{dhakar2022electroviscous} for convenience).

\noindent The Poisson's equation describes the total potential ($U$) and charge density ($n^\ast$) relation \citep{davidson2007electroviscous,dhakar2022slip,dhakar2022electroviscous,dhakar2023cfd} and it is expressed as follows.
\begin{gather}
	\nabla^2U=-\frac{1}{2}K^2n^\ast
	\label{eq:2}
\end{gather}
where $n^\ast(=n_+-n_-)$ and $n_\text{j}$ are the excess charge and number density of $j^{th}$ ion, respectively.

\noindent In electroviscous flows (EVFs), total potential ($U$) for homogeneously charged ($S_\text{rh}=1$) uniform geometries is described as follows.
\begin{gather}
	U(x,y)=\psi(y)-xE_\text{x}
	\label{eq:1a}
\end{gather}
where $xE_\text{x}$, $\psi$, $E_\text{x}$, $x$ and $y$ are the streaming, EDL potential, axial induced electric field strength, axial and radial coordinates, respectively. However, for non-uniform and heterogeneously charged ($S_\text{rh}\neq1$) devices, \eqn\ref{eq:1a} is not applicable. Therefore, decoupling of two potentials (i.e., streaming and EDL potentials) is not possible and has to considered the total potential for present numerical study.

\noindent The Poisson's equation (i.e., \eqn\ref{eq:2}) is subjected to the following boundary conditions:

\noindent Uniform potential gradient is used at the inflow ($x=0$) and outflow ($x=L$) boundary of the channel. It is obtained by `zero current continuity' relation (\eqn\ref{eq:7}) \citep{davidson2007electroviscous,dhakar2022slip,dhakar2022electroviscous,dhakar2023cfd} and it is given below.
\begin{gather}
	I_{\text{net}} = \underbrace{\int_{-1}^{1} {n^\ast\myvec{V}} dy} _{I_{\text{s}}} + \underbrace{\int_{-1}^{1} -{Pe^{-1}\left[\frac{\partial n_{\text{+}}}{\partial x}-\frac{\partial n_{\text{-}}}{\partial x}\right]} dy}_{I_{\text{d}}} + \underbrace{\int_{-1}^{1} -{Pe^{-1}\left[(n_{\text{+}}+n_{\text{-}})\frac{\partial U}{\partial x}\right]} dy}_{I_{\text{c}}} =0
	\label{eq:7}
\end{gather}
where $I_\text{c}$, $I_\text{d}$, $I_\text{s}$, and $\myvec{V}$ are the conduction, diffusion, streaming currents, and velocity vector (defined in \eqn\ref{eq:4}), respectively. At steady-state condition, diffusion current is zero, i.e., $I_\text{d}=0$.

\noindent Charge-heterogeneity (i.e, axial variation of surface charge) is considered at the walls of microfluidic device and it is given below.
\begin{gather}
	(\nabla U\cdot\myvec{n}_{b}) = 
	\begin{cases}	
		S_\text{1} \quad \text{for} \quad a_1 \le x < a_2  \quad \text{and} \quad a_3 < x \le a_4 
		\\ S_\text{2} \quad \text{for} \quad a_2 < x < a_3 
	\end{cases}
	\label{eq:11}
	\\
	\nonumber
	%
	\\
	\text{where,}\qquad
	S_{\text{i}}=\frac{\sigma_{\text{i}}W}{\varepsilon_{\text{0}}\varepsilon_\text{r}U_\text{c}}, \qquad
	i=1, 2
	\nonumber
\end{gather}
where $\myvec{n}_{b}$ and $S$ are the unit normal vector and surface charge density, respectively.

\noindent Further, surface charge-heterogeneity ratio is defined as given below.
\begin{gather}
	S_{\text{rh}}=\frac{S_{\text{2}}}{S_{\text{1}}}
	\label{eq:12}
\end{gather}
Here, $S_1>0$ whereas $S_2\ge 0$, i.e., in case of $S_\text{rh}=0$, only upstream/downstream section walls are charged and contraction region walls are electrically neutral (i.e., $S_2=0$); each section of channel is homogeneously charged for $S_\text{rh}=1$. The upstream/downstream regions walls charge ($S_1$) dominates for $S_\text{rh}<1$ and contraction section walls charge ($S_2$) dominates for $S_\text{rh}>1$.

\noindent Nernst-Planck (NP) equation describes the mass conservation of each ion species ($n_\text{j}$) \citep{davidson2007electroviscous,dhakar2022slip,dhakar2022electroviscous,dhakar2023cfd} and it is given below.
\begin{gather}
	\left[\frac{\partial n_{\text{j}}}{\partial t}+\nabla\cdot(\myvec{V}n_{\text{j}})\right]=\frac{1}{Pe}\left[\nabla^2n_{\text{j}}\pm\nabla\cdot(n_{\text{j}}\nabla U)\right]
	\label{eq:3}
\end{gather}
where $t$ is time.

\noindent The  Nernst-Planck (NP) equation (i.e., \eqn\ref{eq:3}) is subjected to the following boundary conditions:

\noindent At inflow ($x=0$) boundary, ion concentration profile is imposed from the steady fully developed electroviscous flow numerical solution of uniform slit \citep{davidson2007electroviscous}. At outflow ($x=L$) boundary, concentration gradient is applied as zero. At device walls, zero ion flux density ($\myvec{f}_{\text{j}}$) is considered normal to the walls. It is expressed as follows. 
\begin{gather}
	n_\pm=\exp[{\mp\psi(y)}], 
\qquad 
	\frac{\partial n_{\text{j}}}{\partial \myvec{n}_{\text{b}}} = 0,
\qquad
\myvec{f}_{\text{j}}\cdot \myvec{n}_{\text{b}}=0
	\label{eq:6} 
\end{gather}
where $\myvec{f}_{\text{j}}$ flux density of ions described by Einstein relation \citep[refer \eqn A.5 in][]{dhakar2022electroviscous}.

\noindent The velocity ($\myvec{V}$) and pressure ($P$) profiles are described by Navier-Stokes (NS) with an extra electrical body force term and continuity equations \citep{davidson2007electroviscous,dhakar2022slip,dhakar2022electroviscous,dhakar2023cfd}. It is expressed as follows.
\begin{gather}
	\left[\frac{\partial \mathbf{V}}{\partial t}+\nabla\cdot(\myvec{V}\myvec{V})\right]=-\nabla P+\frac{1}{Re}\nabla \cdot\left[\nabla\myvec{V}+(\nabla\myvec{V})^T\right]-\underbrace{\beta\left(\frac{K}{Re}\right)^2(n_{\text{+}}-n_{\text{-}})\nabla U}_{\myvec{F}_{\text{e}}}
	\label{eq:4}
	\\
	\nabla\cdot\myvec{V}=0 \label{eq:5}
\end{gather}
where $P$ and $\myvec{F}_{\text{e}}$ are the pressure and electrical body force, respectively.

\noindent The Navier-Stokes (NS) and continuity equations (i.e., \eqns\ref{eq:4} and \ref{eq:5}) are subjected to the following boundary conditions:

\noindent Fully developed velocity profile is applied at the inflow ($x=0$) boundary from the steady fully developed electroviscous flow of uniform slit numerical solution \citep{davidson2007electroviscous}. It is shown as follows.
\begin{gather}
	V_{\text{x}}=V_{\text{0}}(y), \qquad
	V_{\text{y}}=0
	\label{eq:6a} 
\end{gather}
\noindent Velocity gradient is considered to be zero at outflow ($x=L$) boundary open to ambient. No-slip condition, i.e., both tangential and normal components are zero is applied at device walls. It is shown as follows. 
\begin{gather}
	\frac{\partial \myvec{V}}{\partial \myvec{n}_{\text{b}}} = 0,
	\qquad
	P =0, 
	\qquad
	V_{\myvec{n}_{\text{b}}} =0,
	\qquad 
	V_{\myvec{t}_{\text{b}}} =0
	\label{eq:8}
\end{gather}
where $V_{\myvec{n}_{\text{b}}}$ and $V_{\myvec{t}_{\text{b}}}$ are the normal and tangential components of velocity, respectively.

\noindent The mathematical model consisting of governing equations and boundary conditions (\eqns\ref{eq:2} to \ref{eq:8}) for present physical problem is solved by using the finite element method (FEM) to obtain total potential ($U$), ion concentration ($n_\pm$), induced electric field strength ($E_\text{x}$), excess charge ($n^\ast$), velocity ($\myvec{V}$), and pressure ($P$) in the contraction-expansion microfluidic device for given ranges of governing parameters ($2\le K\le 20$, $4\le S_\text{1}\le 16$, $0\le S_\text{rh}\le 2$). 
\section{Numerical methodology}
\label{sec:sanp}
\noindent The detailed numerical methodology, domain, and mesh independence tests are carried out and presented elsewhere \citep{dhakar2022electroviscous,dhakar2023cfd}. The mathematical model consisting of governing equations with relevant boundary conditions (\eqns\ref{eq:2} to \ref{eq:8}) for electroviscous flow through contraction-expansion heterogeneously charged microfluidic device is solved numerically by using the finite element method (FEM) based COMSOL multiphysics software. Present physical problem can be expressed by COMSOL modules such as \textit{electrostatic} (es), \textit{transport of diluted species} (tds), and \textit{laminar flow} (spf) for total potential, ion concentration, velocity, and pressure profiles, respectively. The \textit{intop} function that in the model coupling defined in the component section is used to solve the integral in the 'zero current continuity equation' (i.e., \eqn\ref{eq:7}). Further, fully coupled PARDISO (PARallel DIrect SOlver) linear and Newton's non-linear solvers are used to obtain the steady-state solution numerically. The steady-state solution yields the total potential ($U$), induced electric field strength ($E_\text{x}$), ion concentrations ($n_\pm$), velocity ($\myvec{V}$), and pressure ($P$) fields. Uniform rectangular mesh with boundary layers and corner refinements is used to discretize the two-dimensional (2-D) computational domain. The $M2$ mesh (100 grid points per half width) is used in the simulations to obtain the present numerical results \citep{dhakar2022electroviscous}.
\section{Results and discussion}
\noindent In this study, pressure-driven flow (PDF) of electrolyte liquid through heterogeneously charged contraction-expansion microfluidic device has been solved to obtain detailed numerical results for wide ranges of dimensionless parameters such as inverse Debye length ($2\le K\le 20$), surface charge density ($4\le S_\text{1}\le 16$), surface charge-heterogeneity ratio ($0\le S_\text{rh}\le 2$), Reynolds number ($Re=10^{-2}$), Schimdt number ($\mathit{Sc}=10^{-3}$), and liquid parameter ($\beta=2.34\times10^{-4}$).

\noindent Before generating the numerical results, the justification of selected parameters ranges is extremely important. The following ranges of parameters: Reynolds number ($Re=10^{-2}$), schimdt number ($\mathit{Sc}=10^{-3}$), inverse Debye length ($K=2$ to 20), surface charge density ($S_1=4$ to 16), surface charge-heterogeneity ratio ($S_\text{rh}=0$ to 2), and liquid parameter ($\beta=2.34\times10^{-4}$) have used in the present study and justified elsewhere \citep{davidson2007electroviscous,dhakar2022slip,dhakar2022electroviscous,dhakar2023cfd}.

\noindent The validation of numerical approach used in the present study is carried out elsewhere \citep{dhakar2022electroviscous} for limiting case of EV flow of electrolyte liquid through homogeneously charged ($S_\text{rh}=1$) contraction-expansion microchannel for wide ranges of conditions; it is thus not repeated herein. Furthermore, none of the results are available for EV flow through heterogeneously charged considered geometry. Based on the previous experience \citep{davidson2007electroviscous,bharti2008steady, bharti2009electroviscous,dhakar2022electroviscous}, present numerical results are reliable and accurate within $\pm$1-2\%. Subsequently, new results are presented in terms of total electrical potential ($U$), induced electric field strength ($E_\text{x}$), ion concentration ($n_\pm$), excess charge ($n^\ast$), pressure ($P$), and electroviscous correction factor ($Y$) as a function of flow governing parameters ($K$, $S_1$, $S_\text{rh}$). 
\subsection{Total potential ($U$)}
\label{sec:potential}
\begin{figure}[t]
	\centering\includegraphics[width=1\linewidth]{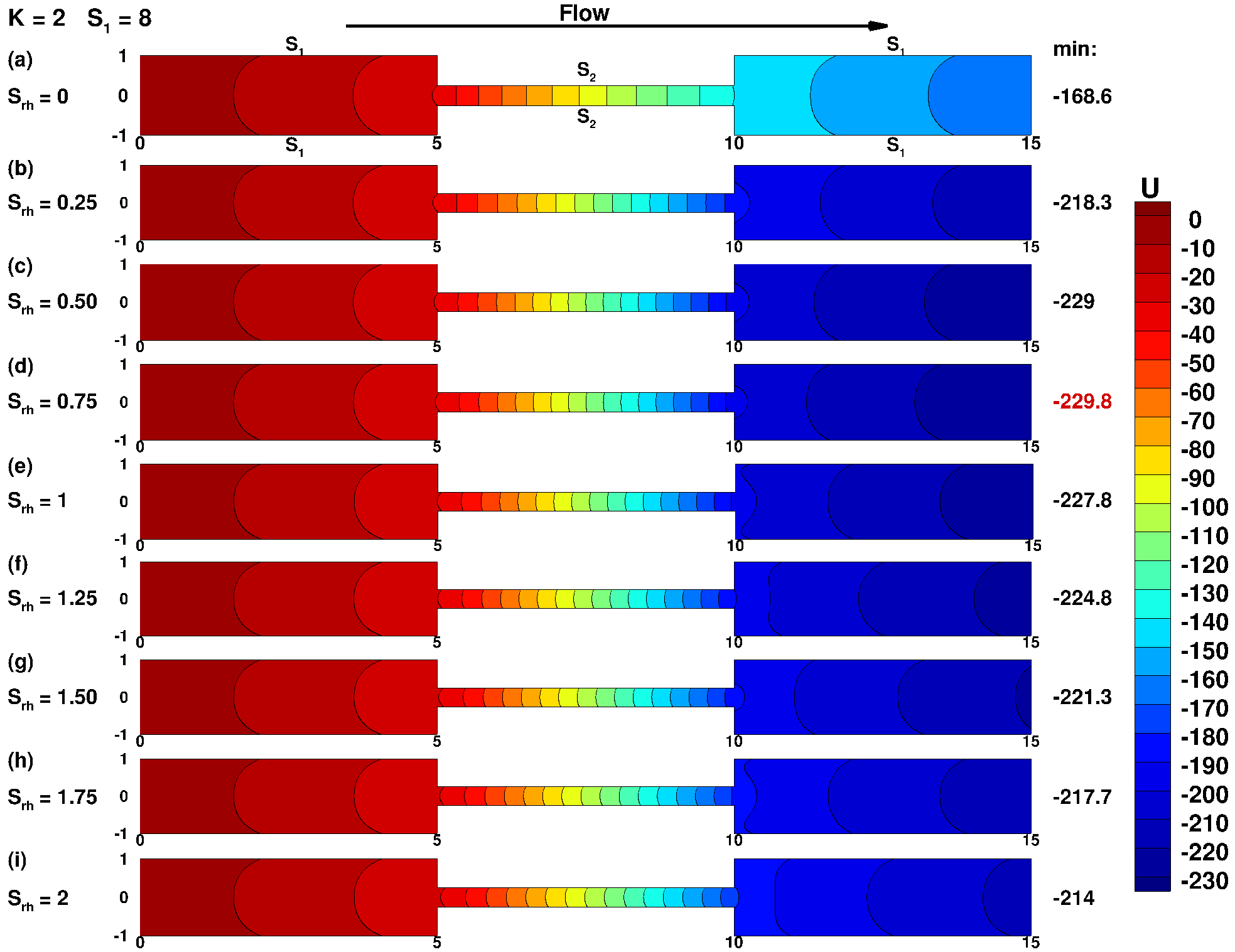}
	\caption{Dimensionless total electrical potential ($U$) distribution for $0\le S_\text{rh}\le 2$, $S_1=8$, and $K=2$.}
	\label{fig:2}
\end{figure}

\noindent \fig\ref{fig:2} depicts the distribution of total electrical potential ($U$) in the considered microfluidic device for $0\le S_\text{rh}\le 2$, $S_1=8$, and $K=2$. Contours of $U$ are qualitatively similar for other ranges of conditions ($2\le K\le 20$, $4\le S_\text{1}\le 16$, $0\le S_\text{rh}\le 2$), thus not shown here. Along the length ($0\le x\le L$) of positively charged channel, $U$ decreases due to advection of excess charge (\fig\ref{fig:2}). Lateral curving in the contours are seen because of $S_\text{i}\neq0$ (i.e., $\partial U/\partial \myvec{n}_{\text{b}}\neq0$) on the device walls (except in contraction section in \fig\ref{fig:2}a). The contours are symmetric about the centreline ($P_0$ to $P_4$; \fig\ref{fig:1}) for homogeneously charged ($S_\text{rh}=1$) condition \citep{davidson2007electroviscous,dhakar2022electroviscous,dhakar2023cfd}. The $U$ decrease with enhancing $S_\text{rh}$, but at higher $S_\text{rh}$, it increases with increasing $S_\text{rh}$ (\fig\ref{fig:2}). Overall minimum value of $U$ is obtained as -229.8 at $S_\text{rh}=0.75$, $K=2$, and $S_1=8$ (\fig\ref{fig:2}d). In general, potential gradient is maximum in the contraction than other regions
(\fig\ref{fig:2}). In addition, shape of contours is significantly affected in the contraction region with the variation of charge-heterogeneity ($0\le S_\text{rh}\le 2$); changes convex to uniform for $S_\text{rh}<1$ followed by enhances convexity for $S_\text{rh}>1$ compare to $S_\text{rh}=1$ (\fig\ref{fig:2}). It is because variation of charge attractive forces in the contraction section with increasing $S_\text{rh}$ (i.e., enhancing $S_\text{2}$ from \eqn\ref{eq:12}).

\begin{figure}[h]
	\centering\includegraphics[width=1\linewidth]{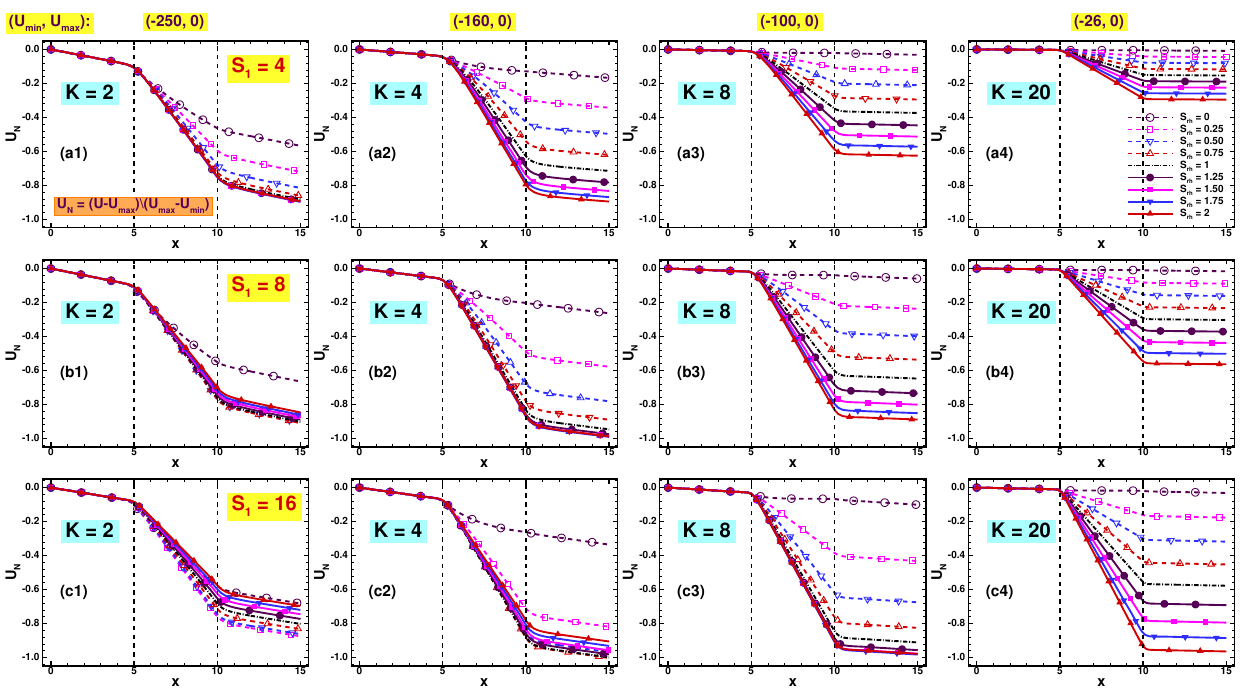}
	\caption{Centreline profiles of normalized total potential ($U_\text{N}$) in heterogeneously charged microfluidic device for $2\le K\le 20$, $4\le S_\text{1}\le 16$, and $0\le S_\text{rh}\le 2$.}
	\label{fig:4a}
\end{figure}

\noindent Further, extensive analysis of total potential is carried out by normalizing the centreline profiles of $U$ with maximum (\textit{max}) and minimum (\textit{min}) values at each $K$; it is defined as $U_\text{N}=(U-U_\text{max})/(U_\text{max}-U_\text{min})$. \fig\ref{fig:4a} shows the centreline ($P_0$ to $P_4$; \fig\ref{fig:1}) profiles of normalized total potential ($U_\text{N}$) in the considered microchannel for $2\le K\le 20$, $4\le S_\text{1}\le 16$, and $0\le S_\text{rh}\le 2$. The $U_\text{N}$ varies in the range of $0\ge U_\text{N}\ge -1$. Centreline profiles of $U_\text{N}$ have depicted the similar qualitative trends as $U$ with the literature \citep{davidson2007electroviscous,dhakar2022electroviscous,dhakar2023cfd} for $S_\text{rh}=1$. The $U_\text{N}$ reduces along the length ($0\le x\le L$) of positively charged device similar to $U$ due to advection of excess charge in the direction of PDF (\fig\ref{fig:4a}). The $U_\text{N}$ increases with increasing $K$ or EDL thinning (\fig\ref{fig:4a}). The $U_\text{N}$ has depicted complex variation with $S_1$ and $S_\text{rh}$. The decrement in $U_\text{N}$ is noted with increasing $S_\text{rh}$ and $S_1$ but opposite trends are obtained at higher $S_1$ and $S_\text{rh}$ (\fig\ref{fig:4a}). In general, $U_{\text{N}}$ has shown maximum value of gradient as $U$ in the contraction than other sections of device. It is because suddenly constricted flow area and increment in charge-heterogeneity ($0\le S_\text{rh}\le 2$) enhances both the clustering of excess charge and velocity in that section. Thus, enhancement in the streaming potential reduces $U$ and increases gradient of $U_\text{N}$ in the contraction than other regions (\fig\ref{fig:4a}).

\noindent Subsequently, \tab\ref{tab:1} summarizes the total electrical potential drop ($|\Delta U|$) on the centreline ($P_0$ to $P_4$; \fig\ref{fig:1}) of considered microfluidic device for $2\le K\le 20$, $4\le S_\text{1}\le 16$, and $0\le {S}_\text{rh}\le 2$. Maximum potential drop ($|\Delta U_\text{max}|$) values are underlined for $0\le S_\text{rh}\le 2$ at each $K$ and $S_1$. The variation in $|\Delta U|$ with $K$ and $S_1$ is same as the literature \citep{davidson2007electroviscous,dhakar2022electroviscous,dhakar2023cfd} for homogeneously charged ($S_\text{rh}=1$) condition. The $|\Delta U|$ decreases with increasing $K$ or thinning of the EDL; maximum variation in $|\Delta U|$ with $K$ is obtained at $S_1=4$ and $S_\text{rh}=0$ (\tab\ref{tab:1}). For instance, $|\Delta U|$ drops with increasing $K$ from 2 to 20 by (99.85\% \add{(141.86 to 0.2127)}, 98.19\% \add{(220.84 to 3.9929)}, 96.56\% \add{(223.52 to 7.6801)}) and (99.52\% \add{(170.56 to 0.8257)}, 92.51\% \add{(200.7 to 15.026)}, 85.61\% \add{(174.27 to 25.084)}) for ($S_\text{rh}=0$, 1, 2), respectively at $S_1=4$ and 16 (refer \tab\ref{tab:1}). The variation in $|\Delta U|$ with $S_1$ is observed maximum at $K=20$ and $S_\text{rh}=0$ (\tab\ref{tab:1}). For instance, $|\Delta U|$ changes with increasing $S_1$ from 4 to 16 by (20.23\% \add{(141.86 to 170.56)}, -9.12\% \add{(220.84 to 200.7)}, -22.03\% \add{(223.52 to 174.27)}) and (288.25\% \add{(0.2127 to 0.8257)}, 276.32\% \add{(3.9929 to 15.026)}, 226.61\% \add{(7.6801 to 25.084)}) for ($S_\text{rh}=0$, 1, 2), respectively at $K=2$ and 20 (refer \tab\ref{tab:1}). The impact of $S_\text{rh}$ on $|\Delta U|$ is obtained maximum at weak-EVF ($S_1=4$ and $K=20$) condition (\tab\ref{tab:1}). For instance, $|\Delta U|$ reduces with decreasing charge-heterogeneity $S_\text{rh}$ from 1 to 0 by (35.76\% \add{(220.84 to 141.86)}, 26.26\% \add{(225.6 to 166.35)}, 15.02\% \add{(200.7 to 170.56)}) and (94.67\% \add{(3.9929 to 0.2127)}, 94.64\% \add{(7.8842 to 0.4227)}, 94.51\% \add{(15.026 to 0.8257)}) for ($S_1=4$, 8, 16), respectively at $K=2$ and 20; on the other hand, the variation in $|\Delta U|$ is noted with increasing charge-heterogeneity $S_\text{rh}$ from 1 to 2 as (1.21\% \add{(220.84 to 223.52)}, -6.12\% \add{(225.6 to 211.8)}, -13.17\% \add{(200.7 to 174.27)}) and (92.34\% \add{(3.9929 to 7.6801)}, 85.94\% \add{(7.8842 to 14.66)}, 66.94\% \add{(15.026 to 25.084)}), respectively at $K=2$ and 20. Overall increment in the values of $|\Delta U|$ are recorded as (57.56\% \add{(141.86 to 223.52)}, 27.32\% \add{(166.35 to 211.8)}, 2.18\% \add{(170.56 to 174.27)}) and (3511.45\% \add{(0.2127 to 7.6801)}, 3368.51\% \add{(0.4227 to 14.66)}, 2938.09\% \add{(0.8257 to 25.084)}) for ($S_1=4$, 8, 16), respectively at $K=2$ and 20 with overall increasing charge-heterogeneity $S_\text{rh}$ from 0 to 2 ($0\le S_\text{rh}\le 2$) (refer \tab\ref{tab:1}). In general, increment in $|\Delta U|$ is noted with increasing both $S_1$ and $S_\text{rh}$ (i.e., enhances $S_\text{2}$), but at higher $S_\text{rh}$ and $S_1$, it decreases with the increment of $S_1$ and $S_\text{rh}$ (\tab\ref{tab:1}). It attributes that strengthening in electrostatic forces near channel walls increases streaming current and hence $|\Delta U|$, but at higher $S_1$ and $S_\text{rh}$, excess charge flow in the liquid is restricted by significantly stronger charge attractive forces, which reduces $|\Delta U|$ with increasing $S_1$ and $S_\text{rh}$ (refer \tab\ref{tab:1}). 
\begin{sidewaystable}	
	\centering
	\caption{Total potential drop ($|\Delta U|$), minimum excess charge ($n^\ast_{\text{min}}$), maximum induced electric field strength ($E_{\text{x,max}}$), and pressure drop ($10^{-5}|\Delta P|$) on the centreline ($P_0$ to $P_4$; \fig\ref{fig:1}) of heterogeneously charged microfluidic device for $2\le K\le 20$, $4\le S_\text{1}\le 16$, and $0\le S_\text{rh}\le 2$.}\label{tab:1}
	\scalebox{0.59}
	{
		\begin{tabular}{|r|r|r|r|r|r|r|r|r|r|r|r|r|r|r|r|r|r|r|r|}
			\hline
			$S_1$	&	$K$	&	\multicolumn{9}{c|}{$|\Delta U|$} &	\multicolumn{9}{c|}{$n^\ast_{\text{min}}$}	\\\cline{3-20}
			&		&	$S_\text{rh}=0$	&	$S_\text{rh}=0.25$	& $S_\text{rh}=0.50$ &	$S_\text{rh}=0.75$	& $S_\text{rh}=1$ & $S_\text{rh}=1.25$ & $S_\text{rh}=1.50$ & $S_\text{rh}=1.75$ & $S_\text{rh}=2$ &	$S_\text{rh}=0$	&	$S_\text{rh}=0.25$	& $S_\text{rh}=0.50$ &	$S_\text{rh}=0.75$	& $S_\text{rh}=1$ & $S_\text{rh}=1.25$ & $S_\text{rh}=1.50$ & $S_\text{rh}=1.75$ & $S_\text{rh}=2$  \\\hline
			0   & 	$\infty$    & 	0	    & 0	        & 0          & 	0	    & 0         & 	0	& 0	&0	        &0 & 	0	    & 0	        & 0          & 	0	    & 0         & 	0	& 0	&0	        &0  \\\hline
			4	&	2	&	141.8600	&	180.1800	&	203.8900	&	215.4000	&	220.8400	&	223.3200	&	\underline{224.2000}	&	224.1500	&	223.5200	&	-1.2506	&	-2.2537	&	-3.7302	&	-5.1473	&	-6.6593	&	-8.0530	&	-9.3399	&	-10.5320	&	\underline{-11.6400}	\\
			&	4	&	26.8410	&	54.8360	&	79.5240	&	99.4160	&	114.4100	&	125.3700	&	133.3300	&	139.1200	&	\underline{143.3300}	&	-0.2490	&	-0.4069	&	-0.7863	&	-1.1666	&	-1.5275	&	-1.8621	&	-2.1703	&	-2.4551	&	\underline{-2.7194}	\\
			&	6	&	7.6278	&	23.3570	&	38.2720	&	51.8140	&	63.6750	&	73.7920	&	82.2700	&	89.2970	&	\underline{95.0850}	&	-0.0782	&	-0.1546	&	-0.3049	&	-0.4508	&	-0.5901	&	-0.7217	&	-0.8451	&	-0.9608	&	\underline{-1.0691}	\\
			&	8	&	3.1429	&	12.1980	&	21.0390	&	29.4860	&	37.4010	&	44.6940	&	51.3190	&	57.2700	&	\underline{62.5680}	&	-0.0328	&	-0.0686	&	-0.1363	&	-0.2026	&	-0.2669	&	-0.3287	&	-0.3878	&	-0.4440	&	\underline{-0.4973}	\\
			&	20	&	0.2127	&	1.1615	&	2.1088	&	3.0531	&	3.9929	&	4.9269	&	5.8536	&	6.7718	&	\underline{7.6801}	&	-0.0007	&	-0.0013	&	-0.0027	&	-0.0040	&	-0.0053	&	-0.0066	&	-0.0079	&	-0.0092	&	\underline{-0.0105}	\\\hline
			8	&	2	&	166.3500	&	216.0200	&	226.7400	&	\underline{227.5900}	&	225.6000	&	222.5500	&	219.0800	&	215.4500	&	211.8000	&	-1.7315	&	-5.0799	&	-7.0254	&	-9.3707	&	-11.6670	&	-13.6620	&	-15.4050	&	-16.9390	&	\underline{-18.2970}	\\
			&	4	&	42.1430	&	92.5260	&	125.0500	&	142.4500	&	151.4900	&	156.0600	&	158.0900	&	\underline{158.6000}	&	158.1600	&	-0.4142	&	-0.8412	&	-1.5585	&	-2.1715	&	-2.7178	&	-3.1926	&	-3.6083	&	-3.9746	&	\underline{-4.2991}	\\
			&	6	&	13.6940	&	43.8190	&	68.6720	&	86.8250	&	99.3190	&	107.7100	&	113.2600	&	116.8300	&	\underline{119.0200}	&	-0.1452	&	-0.3054	&	-0.5896	&	-0.8433	&	-1.0663	&	-1.2624	&	-1.4350	&	-1.5877	&	\underline{-1.7233}	\\
			&	8	&	5.9502	&	23.6920	&	39.8920	&	39.8920	&	53.6620	&	73.4890	&	80.1530	&	85.1810	&	\underline{88.9160}	&	-0.0635	&	-0.1359	&	-0.2661	&	-0.3866	&	-0.4959	&	-0.5940	&	-0.6817	&	-0.7601	&	\underline{-0.8303}	\\
			&	20	&	0.4227	&	2.3176	&	4.2002	&	6.0594	&	7.8842	&	9.6647	&	11.3920	&	13.0590	&	\underline{14.6600}	&	-0.0013	&	-0.0027	&	-0.0053	&	-0.0079	&	-0.0105	&	-0.0131	&	-0.0156	&	-0.0180	&	\underline{-0.0204}	\\\hline
			16	&	2	&	170.5600	&	\underline{218.9000}	&	215.6600	&	208.3800	&	200.7000	&	193.3400	&	186.4600	&	180.1100	&	174.2700	&	-3.1877	&	-9.1717	&	-11.9740	&	-15.4220	&	-18.3160	&	-20.6070	&	-22.4590	&	-23.9830	&	\underline{-25.2560}	\\
			&	4	&	53.7030	&	131.1100	&	155.0200	&	\underline{160.5100}	&	159.9600	&	157.0900	&	153.2800	&	149.1500	&	144.9900	&	-0.6020	&	-1.7443	&	-2.7544	&	-3.6026	&	-4.2925	&	-4.8400	&	-5.2829	&	-5.6474	&	\underline{-5.9517}	\\
			&	6	&	20.8850	&	73.8330	&	103.2800	&	116.6400	&	122.0800	&	\underline{123.5800}	&	123.0300	&	121.3800	&	119.1500	&	-0.2441	&	-0.5956	&	-1.0642	&	-1.4270	&	-1.7145	&	-1.9435	&	-2.1292	&	-2.2821	&	\underline{-2.4098}	\\
			&	8	&	10.0810	&	43.2320	&	67.5490	&	82.5860	&	91.1610	&	95.7040	&	97.7730	&	\underline{98.3150}	&	97.9060	&	-0.1147	&	-0.2638	&	-0.4923	&	-0.6769	&	-0.8247	&	-0.9440	&	-1.0413	&	-1.1216	&	\underline{-1.1888}	\\
			&	20	&	0.8257	&	4.5946	&	8.2692	&	11.7680	&	15.0260	&	18.0000	&	20.6690	&	23.0270	&	\underline{25.0840}	&	-0.0026	&	-0.0053	&	-0.0105	&	-0.0155	&	-0.0203	&	-0.0248	&	-0.0290	&	-0.0328	&	\underline{-0.0364}	\\\hline
			$S_1$	&	$K$	&	\multicolumn{9}{c|}{$E_\text{x,max}$} &	\multicolumn{9}{c|}{$10^{-5}|\Delta P|$}	\\\cline{3-20}
			&		&	$S_\text{rh}=0$	&	$S_\text{rh}=0.25$	& $S_\text{rh}=0.50$ &	$S_\text{rh}=0.75$	& $S_\text{rh}=1$ & $S_\text{rh}=1.25$ & $S_\text{rh}=1.50$ & $S_\text{rh}=1.75$ & $S_\text{rh}=2$ &	$S_\text{rh}=0$	&	$S_\text{rh}=0.25$	& $S_\text{rh}=0.50$ &	$S_\text{rh}=0.75$	& $S_\text{rh}=1$ & $S_\text{rh}=1.25$ & $S_\text{rh}=1.50$ & $S_\text{rh}=1.75$ & $S_\text{rh}=2$  \\\hline
			0   & 	$\infty$    & 	0	    & 0	        & 0          & 	0	    & 0         & 	0	& 0	&0	        & 0 & 	1.0616	    & 1.0616    & 1.0616          & 	1.0616	    & 1.0616    & 	1.0616	& 1.0616	& 1.0616    & 1.0616  \\\hline
		4	&	2	&	21.4780	&	25.4450	&	31.0290	&	32.9300	&	33.6690	&	33.9560	&	\underline{34.0060}	&	33.9110	&	33.7190	&	1.0762	&	1.0898	&	1.1130	&	1.1402	&	1.1672	&	1.1931	&	1.2176	&	1.2407	&	\underline{1.2625}	\\
		&	4	&	5.0430	&	8.4591	&	14.6670	&	19.1640	&	22.1050	&	23.9810	&	25.2050	&	26.0200	&	\underline{26.5630}	&	1.0637	&	1.0694	&	1.0814	&	1.0985	&	1.1189	&	1.1407	&	1.1628	&	1.1845	&	\underline{1.2056}	\\
		&	6	&	1.5875	&	3.7928	&	7.2790	&	10.3380	&	12.8980	&	14.9780	&	16.6390	&	17.9570	&	\underline{19.0000}	&	1.0621	&	1.0648	&	1.0716	&	1.0820	&	1.0952	&	1.1104	&	1.1268	&	1.1439	&	\underline{1.1612}	\\
		&	8	&	0.6245	&	1.9963	&	3.9349	&	5.7701	&	7.4671	&	9.0062	&	10.3800	&	11.5930	&	\underline{12.6530}	&	1.0617	&	1.0632	&	1.0670	&	1.0731	&	1.0811	&	1.0907	&	1.1017	&	1.1136	&	\underline{1.1262}	\\
		&	20	&	0.0443	&	0.1918	&	0.3827	&	0.5733	&	0.7630	&	0.9515	&	1.1385	&	1.3237	&	\underline{1.5069}	&	1.0616	&	1.0617	&	1.0619	&	1.0624	&	1.0629	&	1.0637	&	1.0646	&	1.0656	&	\underline{1.0668}	\\\hline
		8	&	2	&	26.2340	&	32.0320	&	33.9800	&	\underline{34.0730}	&	33.6930	&	33.0840	&	32.3600	&	31.5790	&	30.7770	&	1.0862	&	1.1288	&	1.1829	&	1.2324	&	1.2766	&	1.3158	&	1.3506	&	1.3814	&	\underline{1.4088}	\\
		&	4	&	8.0877	&	14.8550	&	22.0520	&	25.1200	&	26.4760	&	27.0420	&	\underline{27.1830}	&	27.0760	&	26.8160	&	1.0675	&	1.0876	&	1.1264	&	1.1707	&	1.2136	&	1.2531	&	1.2888	&	1.3209	&	\underline{1.3495}	\\
		&	6	&	2.9023	&	7.2240	&	12.7840	&	16.5100	&	18.8750	&	20.3580	&	21.2680	&	21.8030	&	\underline{22.0810}	&	1.0632	&	1.0736	&	1.0977	&	1.1295	&	1.1640	&	1.1983	&	1.2308	&	1.2611	&	\underline{1.2890}	\\
		&	8	&	1.1964	&	3.9075	&	7.4172	&	10.3170	&	12.5840	&	14.2990	&	15.5710	&	16.4990	&	\underline{17.1610}	&	1.0622	&	1.0678	&	1.0820	&	1.1028	&	1.1274	&	1.1537	&	1.1804	&	1.2064	&	\underline{1.2313}	\\
		&	20	&	0.0841	&	0.3829	&	0.7625	&	1.1378	&	1.5060	&	1.8651	&	2.2133	&	2.5491	&	\underline{2.8713}	&	1.0616	&	1.0620	&	1.0630	&	1.0646	&	1.0669	&	1.0698	&	1.0731	&	1.0770	&	\underline{1.0813}	\\\hline
		16	&	2	&	28.8720	&	\underline{34.2240}	&	33.5160	&	32.1210	&	30.5800	&	29.0360	&	27.5670	&	26.2030	&	24.9490	&	1.0941	&	1.1951	&	1.2875	&	1.3605	&	1.4179	&	1.4628	&	1.4980	&	1.5256	&	\underline{1.5471}	\\
		&	4	&	11.0910	&	21.9020	&	26.3150	&	\underline{27.0300}	&	26.6800	&	25.9280	&	25.0240	&	24.0770	&	23.1410	&	1.0730	&	1.1357	&	1.2229	&	1.2976	&	1.3577	&	1.4054	&	1.4433	&	1.4734	&	\underline{1.4972}	\\
		&	6	&	4.7034	&	12.5180	&	18.5940	&	21.0290	&	21.8790	&	\underline{21.9860}	&	21.7200	&	21.2620	&	20.7080	&	1.0659	&	1.1019	&	1.1683	&	1.2349	&	1.2928	&	1.3409	&	1.3802	&	1.4123	&	\underline{1.4383}	\\
		&	8	&	2.1122	&	7.2797	&	12.3940	&	15.3820	&	16.9880	&	17.7730	&	18.0720	&	\underline{18.0780}	&	17.9050	&	1.0635	&	1.0840	&	1.1294	&	1.1822	&	1.2329	&	1.2778	&	1.3163	&	1.3488	&	\underline{1.3760}	\\
		&	20	&	0.1399	&	0.7617	&	1.5028	&	2.2088	&	2.8655	&	3.4642	&	4.0003	&	4.4729	&	\underline{4.8839}	&	1.0617	&	1.0631	&	1.0671	&	1.0733	&	1.0815	&	1.0912	&	1.1019	&	1.1133	&	\underline{1.1250}	\\\hline
		\end{tabular}
	}
\end{sidewaystable}
\begin{figure}[h]
	\centering\includegraphics[width=1\linewidth]{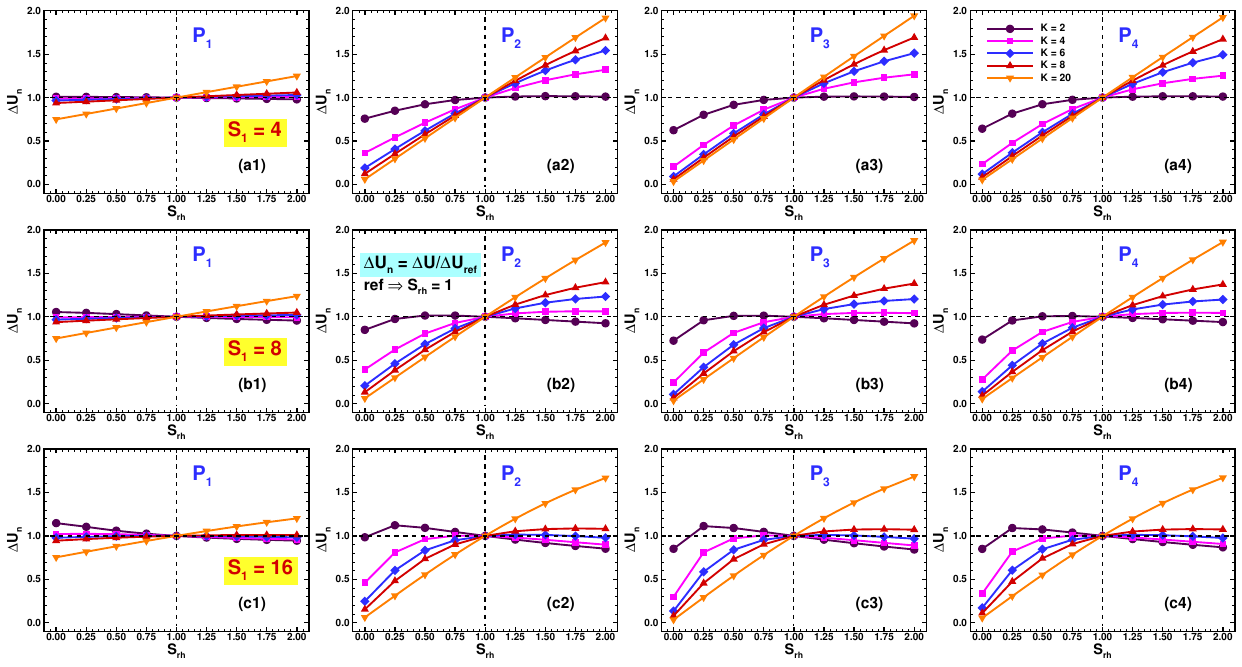}
	\caption{Normalized total potential drop ($\Delta U_\text{n}$) variation on centreline locations ($P_1$, $P_2$, $P_3$, $P_4$; \fig\ref{fig:1}) of heterogeneously charged microfluidic device for $0\le S_\text{rh}\le 2$, $2\le K\le 20$, and $4\le S_\text{1}\le 16$.}
	\label{fig:4}
\end{figure} 

\noindent Further, the relative impact of charge-heterogeneity ($S_\text{rh}$) is described by normalizing the flow fields for heterogeneously charged ($S_\text{rh}\neq1$) by that at reference case of homogeneously charged (`ref' or $S_\text{rh}=1$) device for dimensionless parameters ($S_1$, $K$) as given below \cite{dhakar2023analysis}.
\begin{gather}
	\Psi_\text{N} = \frac{\Psi}{\Psi_{\text{ref}}} = 
	\left.\frac{\Psi(S_\text{rh})}{\Psi(S_\text{rh}=1)}\right|_{S_\text{1}, K}
	\qquad\text{where}\qquad \Psi = (\Delta U, n^{\ast}, E_{\text{x}}, \Delta P)
	\label{eq:11a}
\end{gather}

\noindent Subsequently, total potential variation in the microfluidic device is analyzed in detailed by normalizing with reference potential (at $S_\text{rh}=1$) (refer \eqn\ref{eq:11a}). \fig\ref{fig:4} depicts the normalized total electrical potential ($\Delta U_\text{n}$) variation with $S_\text{rh}$ on centreline locations ($P_1$, $P_2$, $P_3$, $P_4$; \fig\ref{fig:1}) of channel for $2\le K\le 20$ and $4\le S_\text{1}\le 16$. The increment in the values of $\Delta U_\text{n}$ are noted with decreasing $K$ for $S_\text{rh}<1$ but opposite trends are observed for $S_\text{rh}>1$ (\fig\ref{fig:4}). The variation in $\Delta U_\text{n}$ with $K$ is maximum at highest $S_1$ and lowest $S_\text{rh}$ (at $P_3$). For instance, $\Delta U_\text{n}$ decreases maximally by (34.37\% \add{(1.1471 to 0.7529)}, 93.74\% \add{(0.9835 to 0.0616)}, 95.9\% \add{(0.8513 to 0.0349)}, 93.53\% \add{(0.8498 to 0.0549)}) for ($P_1$, $P_2$, $P_3$, $P_4$), respectively when $K$ varies from 2 to 20 at $S_1=16$ and $S_\text{rh}=0$ (refer \fig\ref{fig:4}c). The $\Delta U_\text{n}$ increases with increasing $S_1$ for $S_\text{rh}<1$, but decrement in $\Delta U_\text{n}$ is noted with increasing $S_1$ for $S_\text{rh}>1$, irrespective of $K$ (\fig\ref{fig:4}). The relative impact of $S_1$ on $\Delta U_\text{n}$ is obtained maximum at lower $K$ and $S_\text{rh}$ (at $P_3$). For instance, $\Delta U_\text{n}$ enhances maximally for ($P_1$, $P_2$, $P_3$, $P_4$) by (3.25\% \add{(0.9913 to 1.0235)}, 49.24\% \add{(0.5423 to 0.8093)}, 77.68\% \add{(0.4558 to 0.8099)}, 71.01\% \add{(0.4793 to 0.8196)}), respectively when $S_1$ varies from 4 to 16 at $K=4$ and $S_\text{rh}=0.25$ (refer \fig\ref{fig:4}). Further, enhancement in $\Delta U_\text{n}$ is noted with increasing $S_\text{rh}$, but reverse trends are observed at lower $K$ or thick EDL (\fig\ref{fig:4}). The relative effect of $S_\text{rh}$ on $\Delta U_\text{n}$ is maximum at weak-EVF condition (at $P_3$). For instance, enhancement in the values of $\Delta U_\text{n}$ are noted for ($P_1$, $P_2$, $P_3$, $P_4$) by (66.58\% \add{(0.7494 to 1.2483)}, 3117.6\% \add{(0.0596 to 1.9177)}, 5673.27\% \add{(0.0336 to 1.9425)}, 3511.45\% \add{(0.0533 to 1.9234)}), respectively when $S_\text{rh}$ changes from 0 to 2 at $K=20$ and $S_1=4$ (refer \fig\ref{fig:4}). Thus, it is observed that $\Delta U_\text{n}$ maximally varies with dimensionless parameters ($K$, $S_1$, $S_\text{rh}$) at $P_3$ than other centreline points ($P_1$, $P_2$, $P_4$). It attributes that $P_3$ is significantly affected by relative variation of contraction and downstream sections charge attractive forces and EDL thickness due to sudden expansion in the cross-section flow area at $P_3$ (\fig\ref{fig:4}).

\noindent \eqn(\ref{eq:2}) (i.e., Poisson's equation) relates the total potential ($U$) and charge density ($n^\ast$). Thus, next section present the excess charge ($n^\ast$) distribution in the considered microfluidic geometry as a function of $K$, $S_1$, and $S_\text{rh}$.
\subsection{Excess charge ($n^\ast$)}
\label{sec:charge}
\begin{figure}[!tb]
	\centering\includegraphics[width=0.9\linewidth]{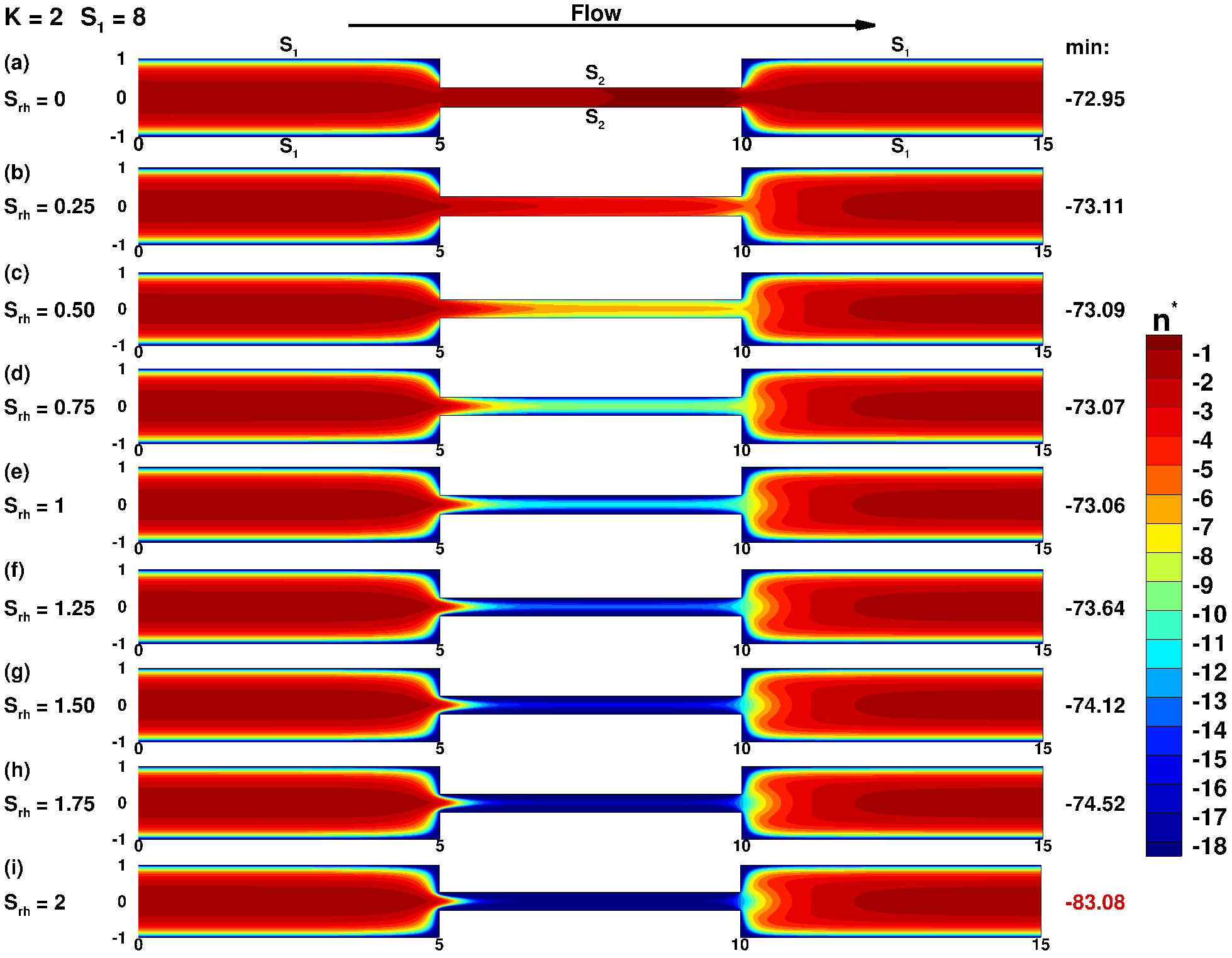}
	\caption{Dimensionless excess charge ($n^\ast$) distribution for $0\le S_\text{rh}\le 2$, $S_1=8$ and $K=2$.}
	\label{fig:5}
\end{figure}

\noindent \fig\ref{fig:5} shows excess charge ($n^\ast$, \eqn\ref{eq:2}) distribution in the considered microfluidic device for $0\le S_\text{rh}\le 2$, $S_\text{1}=8$, and $K=2$. Contours of $n^\ast$ are qualitatively similar for other ranges of conditions ($2\le K\le 20$, $4\le S_\text{1}\le 16$, $0\le S_\text{rh}\le 2$), thus not presented here. The dense clustering of charge is obtained in the close vicinity of the walls in symmetric manner about centreline ($P_0$ to $P_4$; \fig\ref{fig:1}) for homogeneously charged ($S_\text{rh}=1$) condition \citep{dhakar2022electroviscous,dhakar2023cfd} (\fig\ref{fig:5}). Clustering of excess charge is further enhanced for $S_\text{rh}>1$ followed by reduced for $S_\text{rh}<1$ than $S_\text{rh}=1$ in the contraction section (\fig\ref{fig:5}). The $n^\ast$ is minimum in the contraction region because reduction in cross-section area and enhancement in $S_2$ (from \eqn\ref{eq:12} with increasing $S_\text{rh}$) enhances excess charge clustering in that section (\fig\ref{fig:5}). Further, $n^\ast_\text{min}$ decreases with increasing $S_\text{rh}$ due to intensified charge attractive forces in the close vicinity of channel walls (\fig\ref{fig:5}). Lowest value of $n^\ast_\text{min}$ is obtained as -83.08 (at $S_\text{rh}=2$ for fixed $S_1=8$ and $K=2$) (\fig\ref{fig:5}i). However, overall lowest value of $n^\ast$ is noted as -292.59 (at $S_\text{rh}=2$, $S_1=16$, $K=2$). 
\begin{figure}[htbp]
	\centering\includegraphics[width=1\linewidth]{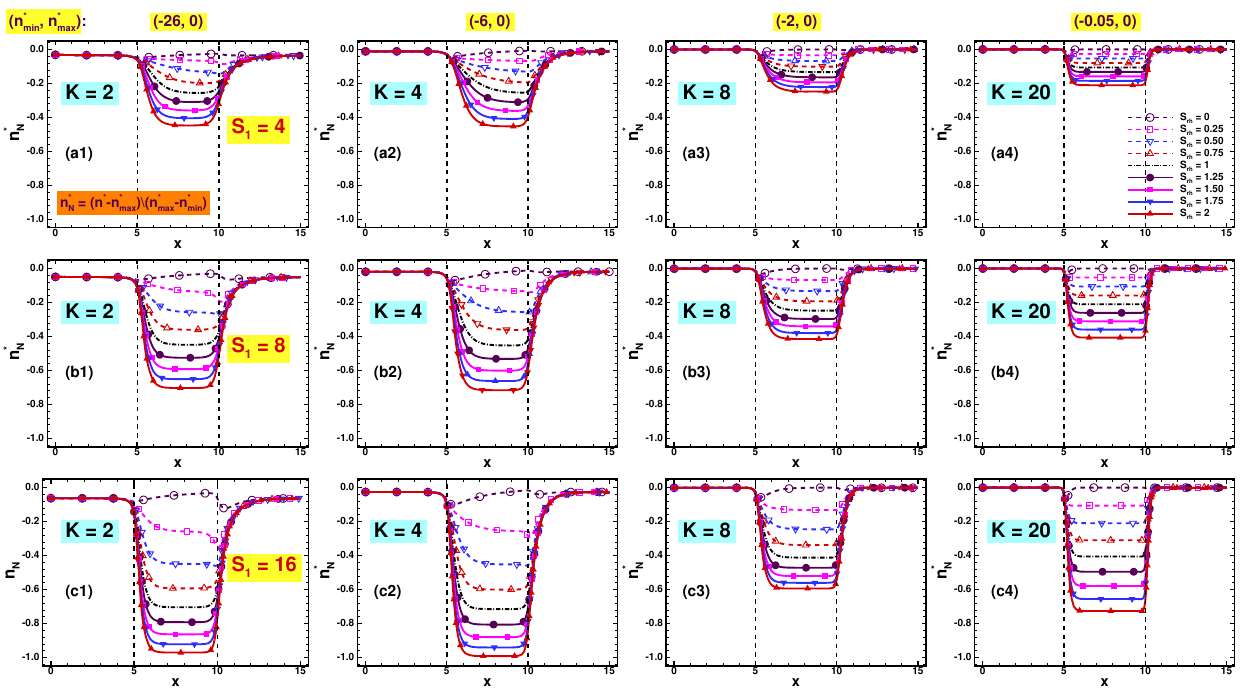}
	\caption{Centreline profiles of normalized excess charge ($n^\ast_\text{N}$) in the heterogeneously charged microfluidic device for $2\le K\le 20$, $4\le S_\text{1}\le 16$, and $0\le S_\text{rh}\le 2$.}
	\label{fig:6a}
\end{figure} 

\noindent Further, extensive analysis of excess charge centreline profiles is carried out by normalizing with minimum (\textit{min}) and maximum (\textit{max}) values of $n^\ast$ at each $K$; it is defined as $n^\ast_\text{N}=(n^\ast-n^\ast_\text{max})/(n^\ast_\text{max}-n^\ast_\text{min})$. \fig\ref{fig:6a} depicts the centreline ($P_0$ to $P_4$; \fig\ref{fig:1}) profiles of normalized excess charge ($n^\ast_\text{N}$) in the considered microfluidic device for $2\le K\le 20$, $4\le S_\text{1}\le 16$, and $0\le S_\text{rh}\le 2$. The $n^\ast_\text{N}$ varies in the range of $0\ge n^\ast_\text{N}\ge -1$. Centreline profiles of $n^\ast_\text{N}$ have depicted similar qualitative behavior as $n^\ast$ with the literature \citep{davidson2007electroviscous,dhakar2022electroviscous,dhakar2023cfd} for $S_\text{rh}=1$. The $n^\ast_\text{N}$ is seen negative ($n^\ast_\text{N}<0$) as $n^\ast$ in the microfluidic device (\fig\ref{fig:6a}). The $n^\ast_\text{N}$ is consistent and equal at the inflow (0, 0) and outflow ($L$, 0) center points of microfluidic device (\fig\ref{fig:6a}). In general, $n^\ast_\text{N}$ is significantly smaller similar to $n^\ast$ in the contraction than other sections of device. It is due to reduction in the cross-section flow area and increment in $S_\text{rh}$ (i.e., increases $S_2$ from \eqn\ref{eq:12}) enhance the excess charge in the that region of device (\fig\ref{fig:6a}). The $n^\ast_\text{N,min}$ has depicted stronger dependency on the dimensionless parameters ($K$, $S_1$, $S_\text{rh}$). The $n^\ast_\text{N,min}$ decreases with decreasing $K$ or thickening of the EDL. Further, $n^\ast_\text{N,min}$ decreases with increasing both $S_1$ and $S_\text{rh}$ (\fig\ref{fig:6a}).

\noindent Subsequently, \tab\ref{tab:1} comprises the minimum excess charge ($n^\ast_\text{min}$) on the centreline ($P_0$ to $P_4$; \fig\ref{fig:1}) of considered microchannel for $2\le K\le 20$, $4\le S_\text{1}\le 16$, and $0\le {S}_\text{rh}\le 2$. Lowest values of $n^\ast_\text{min}$ are underlined for $0\le S_\text{rh}\le 2$ at each $S_1$ and $K$. The variation in $n^\ast_\text{min}$ with $K$ and $S_1$ is same as the literature \citep{davidson2007electroviscous,dhakar2022electroviscous,dhakar2023cfd} for homogeneously charged ($S_\text{rh}=1$) condition. The $n^\ast_\text{min}$ increases with increasing $K$ or thinning of EDL; $n^\ast_\text{min}$ tends to zero when $K\rightarrow \infty$. The maximum variation in $n^\ast_\text{min}$ with $K$ is obtained at $S_1=4$ and $S_\text{rh}=0$ (\tab\ref{tab:1}). For instance, $n^\ast_\text{min}$ reduces with increasing $K$ from 2 to 20 by (99.95\% \add{(-1.2506 to -0.0007)}, 99.92\% \add{(-6.6593 to -0.0053)}, 99.91\% \add{(-11.64 to -0.0105)}) and (99.92\% \add{(-3.1877 to -0.0026)}, 99.89\% \add{(-18.316 to -0.0203)}, 99.86\% \add{(-25.256 to -0.0364)}) for ($S_\text{rh}=0$, 1, 2), respectively at $S_1=4$ and 16 (refer \tab\ref{tab:1}). The effect of $S_1$ on $n^\ast_\text{min}$ is maximum at $K=20$ and $S_\text{rh}=0$ (\tab\ref{tab:1}). For instance, increment in the values of $n^\ast_\text{min}$ are noted for ($S_\text{rh}=0$, 1, 2) as (154.89\% \add{(-1.2506 to -3.1877)}, 175.04\% \add{(-6.6593 to -18.316)}, 116.98\% \add{(-11.64 to -25.256)}) and (292.15\% \add{(-0.0007 to -0.0026)}, 282.11\% \add{(-0.0053 to -0.0203)}, 245.37\% \add{(-0.0105 to -0.0364)}), respectively at $K=2$ and 20 with increasing $S_1$ from 4 to 16 (refer \tab\ref{tab:1}). The change in $n^\ast_\text{min}$ with $S_\text{rh}$ is observed maximum at weak-EVF ($S_1=4$, $K=20$) condition (\tab\ref{tab:1}). For instance, $n^\ast_\text{min}$ decreases with decreasing charge-heterogeneity $S_\text{rh}$ from 1 to 0 by (81.22\% \add{(-6.6593 to -1.2506)}, 85.16\% \add{(-11.667 to -1.7315)}, 82.60\% \add{(-18.316 to -3.1877)}) and (87.68\% \add{(-0.0053 to -0.0007)}, 87.61\% \add{(-0.0105 to -0.0013)}, 87.35\% \add{(-0.0203 to -0.0026)}) for ($S_1=4$, 8, 16), respectively at $K=2$ and 20; on the other hand, enhancement in the values of $n^\ast_\text{min}$ are obtained with increasing charge-heterogeneity $S_\text{rh}$ from 1 to 2 as (74.79\% \add{(-6.6593 to -11.64)}, 56.83\% \add{(-11.667 to -18.297)}, 37.89\% \add{(-18.316 to -25.256)}) and (98.27\% \add{(-0.0053 to -0.0105)}, 93.50\% \add{(-0.0105 to -0.0204)}, 79.20\% \add{(-0.0203 to -0.0364)}), respectively at $K=2$ and 20. Overall increment in $n^\ast_\text{min}$ is noted as (830.75\% \add{(-1.2506 to -11.64)}, 956.71\% \add{(-1.7315 to -18.297)}, 692.30\% \add{(-3.1877 to -25.256)}) and (1508.89\% \add{(-0.0007 to -0.0105)}, 1461.64\% \add{(-0.0013 to -0.0204)}, 1316.93\% \add{(-0.0026 to -0.0364)}) for ($S_1=4$, 8, 16), respectively at $K=2$ and 20 with overall increasing charge-heterogeneity $S_\text{rh}$ from 0 to 2 ($0\le S_\text{rh}\le 2$) (refer \tab\ref{tab:1}). In summary, decrement in the values of $n^\ast_\text{min}$ are noted with increasing both $S_1$ and $S_\text{rh}$. It is because intensified charge attractive forces in the close vicinity of channel walls enhance clustering of excess charge in the EDL, thus, $n^\ast_\text{min}$ decreases with increasing $S_1$ and $S_\text{rh}$ (refer \tab\ref{tab:1}). 
\begin{figure}[h]
	\centering\includegraphics[width=1\linewidth]{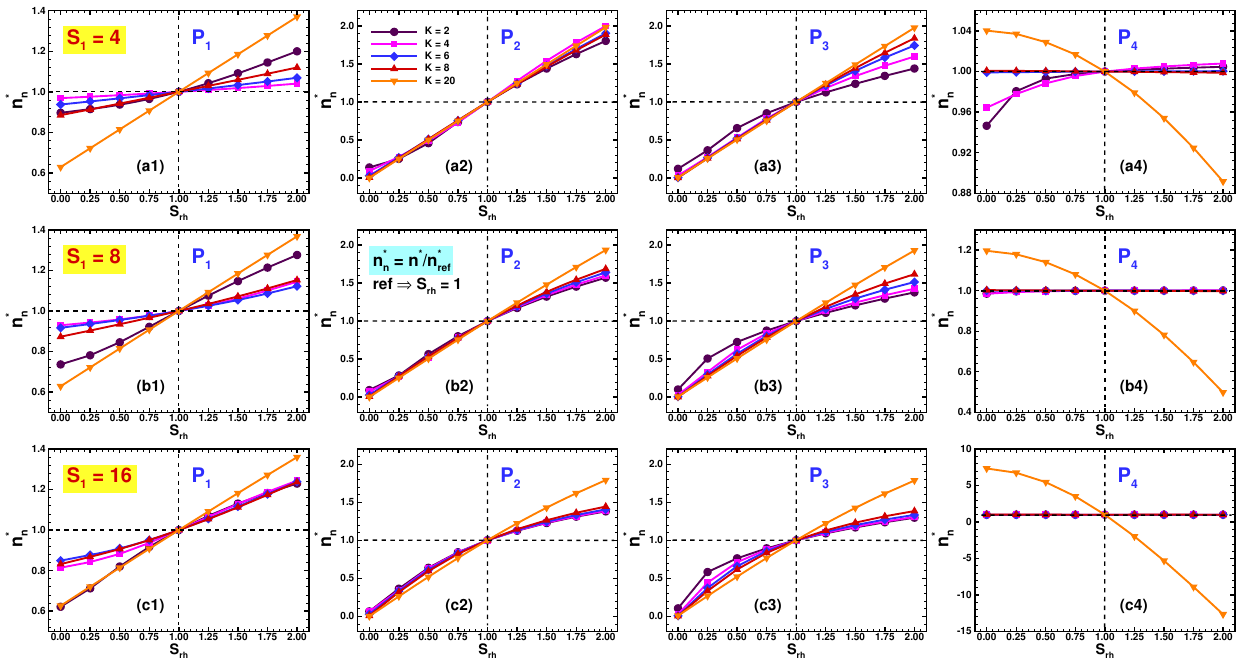}
	\caption{Normalized excess charge ($n^\ast_\text{n}$) variation on centreline locations ($P_1$, $P_2$, $P_3$, $P_4$; \fig\ref{fig:1}) of the heterogeneously charged microfluidic device for $0\le S_\text{rh}\le 2$, $2\le K\le 20$, and $4\le S_\text{1}\le 16$.}
	\label{fig:7}
\end{figure} 

\noindent Further, excess charge variation in the considered microchannel is analyzed in detailed by normalizing with reference ($S_\text{rh}=1$) condition (refer \eqn\ref{eq:11a}). \fig\ref{fig:7} depicts the normalized excess charge ($n^\ast_\text{n}$) variation with $S_\text{rh}$ on centreline locations ($P_1$, $P_2$, $P_3$, $P_4$; \fig\ref{fig:1}) of device for $2\le K\le 20$ and $4\le S_\text{1}\le 16$. The $n^\ast_\text{n}$ increases with decreasing $K$ for $S_\text{rh}<1$, but it reduces with increasing $K$ for $S_\text{rh}>1$ for all centreline points of device. The effect of $K$ on $n^\ast_\text{n}$ is maximum at lowest $S_1$ and $S_\text{rh}$ (at $P_2$) (\fig\ref{fig:7}). For instance, variation in the values of $n^\ast_\text{n}$ are maximally noted as (29.88\% \add{(0.8965 to 0.6286)}, 100\% \add{(0.1386 to 2.2927$\times10^{-8}$)}, 94.57\% \add{(0.1206 to 0.0066)}, 9.92\% \add{(0.9465 to 1.0404)}) for ($P_1$, $P_2$, $P_3$, $P_4$), respectively with increasing $K$ from 2 to 20 at $S_1=4$ and $S_\text{rh}=0$ (refer \fig\ref{fig:7}). The $n^\ast_\text{n}$ increases with increasing $S_1$; maximum variation in $n^\ast_\text{n}$ with $S_1$ is noted at lower $K$ and $S_\text{rh}$ (at $P_3$) (\fig\ref{fig:7}). For instance, $n^\ast_\text{n}$ varies maximally for ($P_1$, $P_2$, $P_3$, $P_4$) by (-13.68\% \add{(0.9760 to 0.8425)}, 22.08\% \add{(0.2780 to 0.3393)}, 60.89\% \add{(0.2742 to 0.4411)}, 2.08\% \add{(0.9778 to 0.9982)}), respectively when $S_1$ varies from 4 to 16 at $K=4$ and $S_\text{rh}=0.25$ (refer \fig\ref{fig:7}). The $n^\ast_\text{n}$ enhances with increasing $S_\text{rh}$ (i.e., increasing $S_2$); the impact of $S_\text{rh}$ on $n^\ast_\text{n}$ is noted maximum at weak-EVF condition (at $P_2$) (\fig\ref{fig:7}). For instance, change in the values of $n^\ast_\text{n}$ are maximally recorded as (117.94\% \add{(0.6286 to 1.3700)}, 8.59$\times10^9$\% \add{(2.2927$\times10^{-8}$ to 1.9825)}, 3.01$\times10^4$\% \add{(0.0066 to 1.9762)}, -14.31\% \add{(1.0404 to 0.8915)}) for ($P_1$, $P_2$, $P_3$, $P_4$), respectively with increasing $S_\text{rh}$  from 0 to 2 at $K=20$ and $S_1=4$ (refer \fig\ref{fig:7}). In summary, maximum variation in $n^\ast_\text{n}$ with flow governing parameters ($K$, $S_1$, $S_\text{rh}$) is obtained at $P_2$ and $P_3$ than other centreline points ($P_1$ and $P_4$). It is because variation in $\Delta U_\text{n}$ is maximum at $P_3$ as discussed in the section \ref{sec:potential} (refer \fig\ref{fig:4}), thus, from \eqn(\ref{eq:2}) maximum variation in $n^\ast$ and hence $n^\ast_\text{n}$ are obtained at $P_2$ and $P_3$ than other centreline points ($P_1$ and $P_4$) (\fig\ref{fig:7}). 
\subsection{Induced electric field strength ($E_{\text{x}}$)}
\label{sec:electric}
\begin{figure}[h]
	\centering\includegraphics[width=1\linewidth]{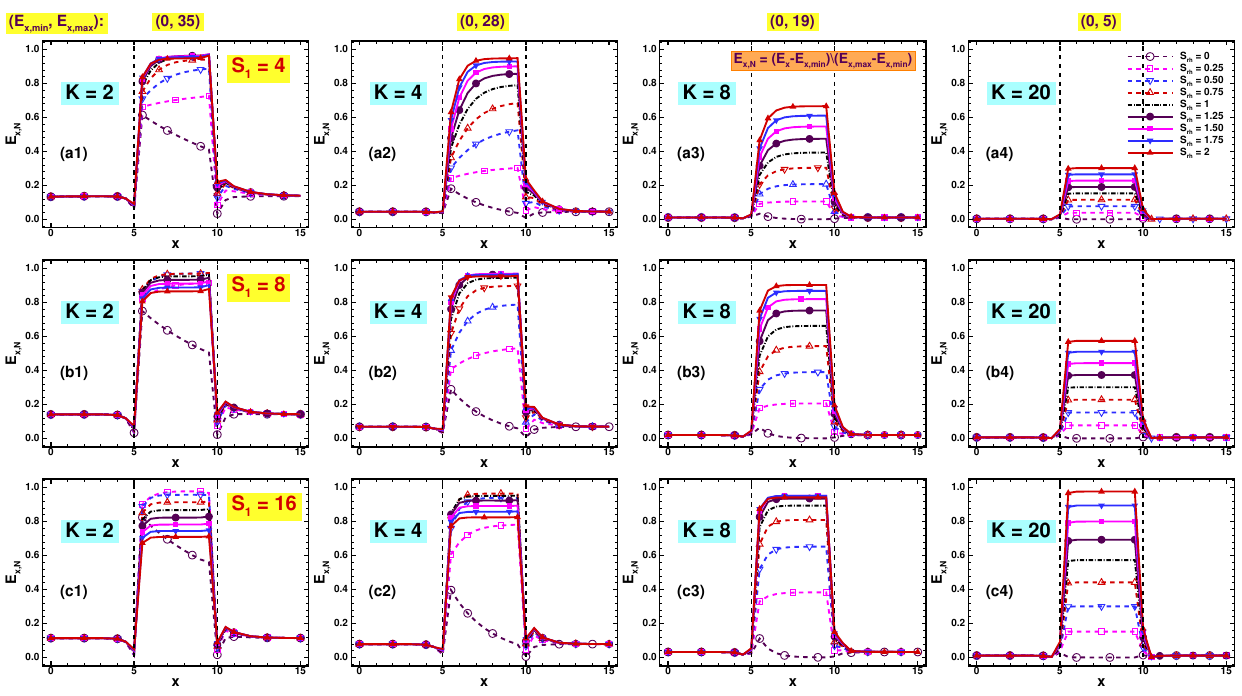}
	\caption{Centreline profiles of normalized induced electric field strength ($E_\text{x,N}$) in the heterogeneously charged microfluidic device for $2\le K\le 20$, $4\le S_\text{1}\le 16$, and $0\le S_\text{rh}\le 2$.}
	\label{fig:8a}
\end{figure}

\noindent The convective flow of excess charge ($n^\ast$) by imposed PDF induces an electric field is called as induced electric field strength ($E_\text{x}=-\partial U/\partial x$), which is calculated numerically from `zero current continuity equation' (i.e., \eqn\ref{eq:7}). The centreline profiles of $E_\text{x}$ are normalized with maximum (\textit{max}) and minimum (\textit{min}) values of $E_\text{x}$ at each $K$ for extensive analysis of $E_\text{x}$ in the common generalized range, over the ranges of flow conditions. It is defined as $E_\text{x,N}=(E_\text{x}-E_\text{x,min})/(E_\text{x,max}-E_\text{x,min})$. \fig\ref{fig:8a} shows centreline ($P_0$ to $P_4$; \fig\ref{fig:1}) profiles of normalized induced electric field strength ($E_\text{x,N}$) in the considered microfluidic device for $2\le K\le 20$, $4\le S_\text{1}\le 16$, and $0\le S_\text{rh}\le 2$. The $E_\text{x,N}$ varies in the range of $0\le E_\text{x,N}\le 1$. Qualitatively similar trends are observed for centreline profiles of $E_\text{x,N}$ as $E_\text{x}$ with the literature \citep{davidson2007electroviscous,dhakar2022electroviscous,dhakar2023cfd} for $S_\text{rh}=1$. In general, $E_\text{x, N}$ is uniform throughout the upstream region ($0\le x\le 5$) except reduces before the contraction ($x\lesssim 5$). In the contraction region ($5\le x\le 10$), $E_\text{x,N}$ depicts such a steep increase along the length followed by smooth enhance with reducing gradient in the middle part and a sudden drop in the end of the section. Further, in downstream region ($10\le x\le 15$), $E_\text{x,N}$ increases initially followed by slow decrement and then become constant in second part of downstream section. In addition, normalized electric field is observed as $E_\text{x,N}\le 0.2$ in upstream/downstream regions, over the ranges of conditions. The $E_\text{x,N}$ has shown significant higher value in contraction than other sections, over the ranges of conditions. It is because suddenly constricted flow area and enhancement in $S_\text{2}$ (i.e., increases $S_\text{rh}$ from \eqn\ref{eq:12}) increase both clustering of excess charge and velocity in that section, thus, increase $E_\text{x}$ and hence $E_\text{x,N}$ in that section (\fig\ref{fig:8a}). The $E_\text{x,N,max}$ decreases with increasing $K$ or thinning of the EDL. Further, $E_\text{x,N,max}$ increases with increasing $S_1$ and $S_\text{rh}$, but decrement in $E_\text{x,N}$ is noted with increasing $S_1$ and $S_\text{rh}$ at higher $S_1$, $S_\text{rh}$ and lower $K$ (\fig\ref{fig:8a}). 

\noindent Subsequently, \tab\ref{tab:1} summarizes the maximum induced electric field strength ($E_\text{x,max}$) on the centreline ($P_0$ to $P_4$; \fig\ref{fig:1}) of considered microchannel for $2\le K\le 20$, $4\le S_\text{1}\le 16$, and $0\le {S}_\text{rh}\le 2$. Highest values of $E_\text{x,max}$ are underlined for $0\le S_\text{rh}\le 2$ at each $S_1$ and $K$. The variation in $E_\text{x,max}$ with $K$ and $S_1$ is same as the literature \citep{dhakar2022electroviscous,dhakar2023cfd} for homogeneously charged ($S_\text{rh}=1$) condition. The $E_\text{x,max}$ increases with decreasing $K$ or thickening of EDL; maximum variation in $E_\text{x,max}$ with $K$ is observed at $S_1=4$ and $S_\text{rh}=0$ (\tab\ref{tab:1}). For instance, $E_\text{x,max}$ reduces with increasing $K$ from 2 to 20 by (99.79\% \add{(21.478 to 0.0443)}, 97.73\% \add{(33.669 to 0.763)}, 95.53\% \add{(33.719 to 1.5069)}) and (99.52\% \add{(28.872 to 0.1399)}, 90.63\% \add{(30.58 to 2.8655)}, 80.42\% \add{(24.949 to 4.8839)}) for ($S_\text{rh}=0$, 1, 2), respectively at $S_1=4$ and 16 (refer \tab\ref{tab:1}). The change in $E_\text{x,max}$ with $S_1$ is maximum at $S_\text{rh}=1$ and $K=20$ (\tab\ref{tab:1}). For instance, change in the values of $E_\text{x,max}$ are recorded as (34.43\% \add{(21.478 to 28.872)}, -9.17\% \add{(33.669 to 30.58)}, -26.01\% \add{(33.719 to 24.949)}) and (216.13\% \add{(0.0443 to 0.1399)}, 275.55\% \add{(0.763 to 2.8655)}, 224.10\% \add{(1.5069 to 4.8839)}) for ($S_\text{rh}=0$, 1, 2), respectively at $K=2$ and 20 with increasing $S_1$ from 4 to 16 (refer \tab\ref{tab:1}). The impact of $S_\text{rh}$ on $E_\text{x,max}$ is obtained maximum at $K=20$ and $S_1=16$ (\tab\ref{tab:1}). For instance, $E_\text{x,max}$ reduces with decreasing charge-heterogeneity $S_\text{rh}$ from 1 to 0 by (36.21\% \add{(33.669 to 21.478)}, 22.14\% \add{(33.693 to 26.234)}, 5.59\% \add{(30.58 to 28.872)}) and (94.20\% \add{(0.763 to 0.0443)}, 94.42\% \add{(1.506 to 0.0841)}, 95.12\% \add{(2.8655 to 0.1399)}) for ($S_1=4$, 8, 16), respectively at $K=2$ and 20; on the other hand, variation in the values of $E_\text{x,max}$ are noted with increasing charge-heterogeneity $S_\text{rh}$ from 1 to 2 as (0.15\% \add{(33.669 to 33.719)}, -8.65\% \add{(33.693 to 30.777)}, -18.41\% \add{(30.58 to 24.949)}) and (97.49\% \add{(0.763 to 1.5069)}, 90.66\% \add{(1.506 to 2.8713)}, 70.44\% \add{(2.8655 to 4.8839)}), respectively at $K=2$ and 20. Overall enhancement in $E_\text{x,max}$ is recorded as (56.99\% \add{(21.478 to 33.719)}, 17.32\% \add{(23.234 to 30.777)}, -13.59\% \add{(28.872 to 24.949)}) and (3304.89\% \add{(0.0443 to 1.5069)}, 3316.14\% \add{(0.0841 to 2.8713)}, 3390.74\% \add{(0.1399 to 4.8839)}) for ($S_1=4$, 8, 16), respectively at $K=2$ and 20 with overall increasing charge-heterogeneity $S_\text{rh}$ from 0 to 2 ($0\le S_\text{rh}\le 2$) (refer \tab\ref{tab:1}). In general, $E_\text{x,max}$ has depicted complex dependence on $S_1$ and $S_\text{rh}$. The enhancement in $E_\text{x,max}$ is noted with increasing both $S_1$ and $S_\text{rh}$, but opposite trends are observed at higher $S_1$ and $S_\text{rh}$. It is because convective $n^\ast$ increases with increasing $S_1$ and $S_\text{rh}$ as discussed in the section \ref{sec:charge}. Thus, $E_\text{x,max}$ enhances with increasing $S_1$ and $S_\text{rh}$. However, at higher $S_1$ and $S_\text{rh}$, charge attractive forces are much stronger, which impedes the convective flow of $n^\ast$ in the device and reduces $E_\text{x,max}$ (refer \tab\ref{tab:1}). 
\begin{figure}[h]
	\centering\includegraphics[width=1\linewidth]{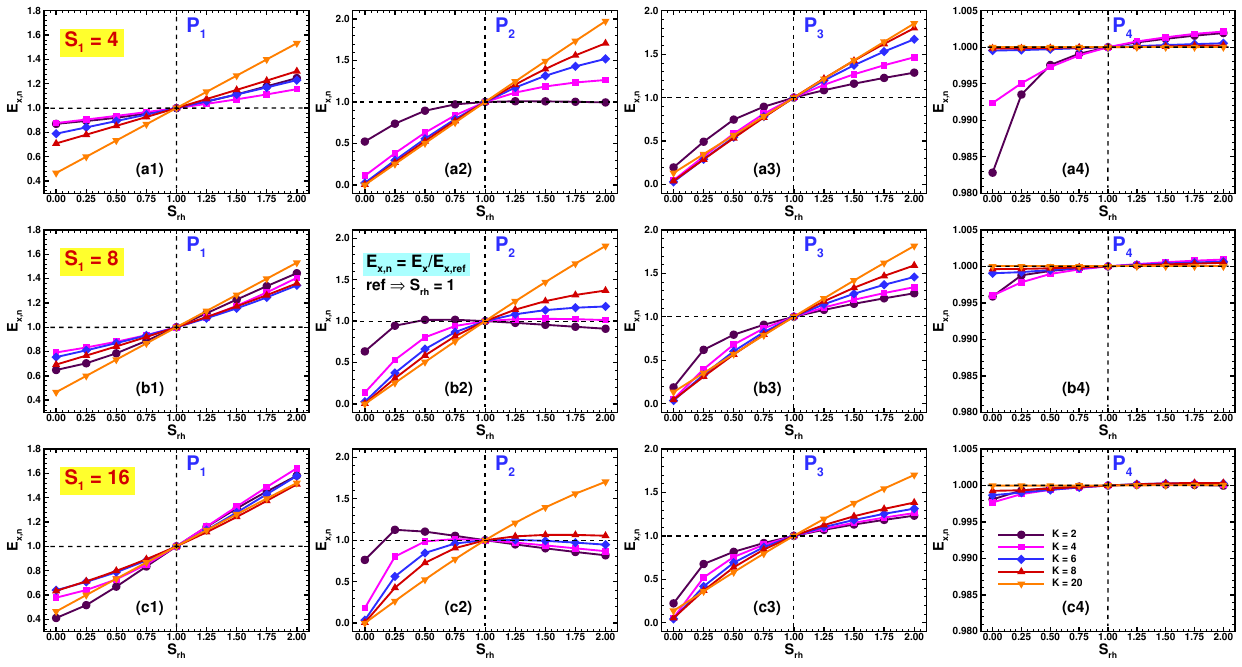}
	\caption{Normalized induced electric field strength ($E_\text{x,n}$) variation on centreline points ($P_1$, $P_2$, $P_3$, $P_4$; \fig\ref{fig:1}) of heterogeneously charged microfluidic device for $0\le S_\text{rh}\le 2$, $4\le S_\text{1}\le 16$, and $2\le K\le 20$.}
	\label{fig:8b}
\end{figure} 

\noindent Further, detailed analysis of induced electric field strength in considered microfluidic device is carried out by normalizing $E_\text{x}$ with reference ($S_\text{rh}=1$) condition (refer \eqn\ref{eq:11a}). \fig\ref{fig:8b} shows normalized induced electric field strength ($E_\text{x,n}$) variation with $S_\text{rh}$ on centreline points ($P_1$, $P_2$, $P_3$, $P_4$) of channel for $2\le K\le 20$ and $4\le S_1\le 16$. The increment in $E_\text{x,n}$ is noted with decreasing $K$ for $S_\text{rh}<1$, but opposite trends are obtained for $S_\text{rh}>1$, irrespective of $S_1$. The impact of $K$ on $E_\text{x,n}$ is maximum at lowest $S_\text{rh}$ and $S_1$ (at $P_2$). For instance, $E_\text{x,n}$ varies maximally by (-46.46\% \add{(0.8701 to 0.4659)}, -100\% \add{(0.5223 to 2.1894$\times10^{-9}$)}, -32.39\% \add{(0.1960 to 0.1327)}, 1.75\% \add{(0.9828 to 1)}) for ($P_1$, $P_2$, $P_3$, $P_4$), respectively when $K$ changes from 2 to 20 at $S_\text{rh}=0$ and $S_1=4$ (refer \fig\ref{fig:8b}). The $E_\text{x,n}$ increases with increasing $S_1$; maximum variation in $E_\text{x,n}$ with $S_1$ is noted at lower $K$ and $S_\text{rh}$ (at $P_2$). For instance, variation in $E_\text{x,n}$ is noted maximally as (-29.29\% \add{(0.9055 to 0.6403)}, 100.26\% \add{(0.3817 to 0.8026)}, 61.83\% \add{(0.3221 to 0.5212)}, 0.38\% \add{(0.9951 to 0.9988)}) for ($P_1$, $P_2$, $P_3$, $P_4$), respectively with increasing $S_1$ from 4 to 16 at $K=4$ and $S_\text{rh}=0.25$ (refer \fig\ref{fig:8b}). The $E_\text{x,n}$ enhances with increasing $S_\text{rh}$, but reverse trends are observed at higher $S_\text{rh}$ and lower $K$. The change in $E_\text{x,n}$ with $S_\text{rh}$ is obtained maximum at weak-EVF condition (at $P_2$). For instance, $E_\text{x,n}$ increases maximally for ($P_1$, $P_2$, $P_3$, $P_4$) by (229.06\% \add{(0.4659 to 1.5331)}, 9.02$\times10^{10}$ \add{(2.1897$\times10^{-9}$ to 1.9749)}, 1.30$\times10^3$ \add{(0.1327 to 1.8537)}, 0\% (1 to 1)), respectively with increasing $S_\text{rh}$ from 0 to 2 at $K=20$ and $S_1=4$ (refer \fig\ref{fig:8b}). In summary, it is noted that variation in $E_\text{x,n}$ with governing parameters ($K$, $S_1$, $S_\text{rh}$) is maximum at $P_2$ than other centreline locations ($P_1$, $P_3$, $P_4$) of device. It is because change in $n^\ast_\text{n}$ is obtained maximum at $P_2$ as discussed in the section \ref{sec:charge} (refer \fig\ref{fig:7}), therefore, maximum change in $E_\text{x,n}$ (from \eqn\ref{eq:7}) is obtained at $P_2$ than other centreline points ($P_1$, $P_3$, $P_4$) of microfluidic device (\fig\ref{fig:8b}). 

\noindent The preceding sections have summarized that total potential ($U$), excess charge ($n^\ast$), and induced electric field strength ($E_\text{x}$) are strongly dependent on the flow governing parameters ($K$, $S_1$, $S_\text{rh}$). Thus, pressure ($P$) has also depends on these parameters and presented in the subsequent section herein. 
\subsection{Pressure ($P$)}
\label{sec:pressure}
\begin{figure}[h]
	\centering\includegraphics[width=1\linewidth]{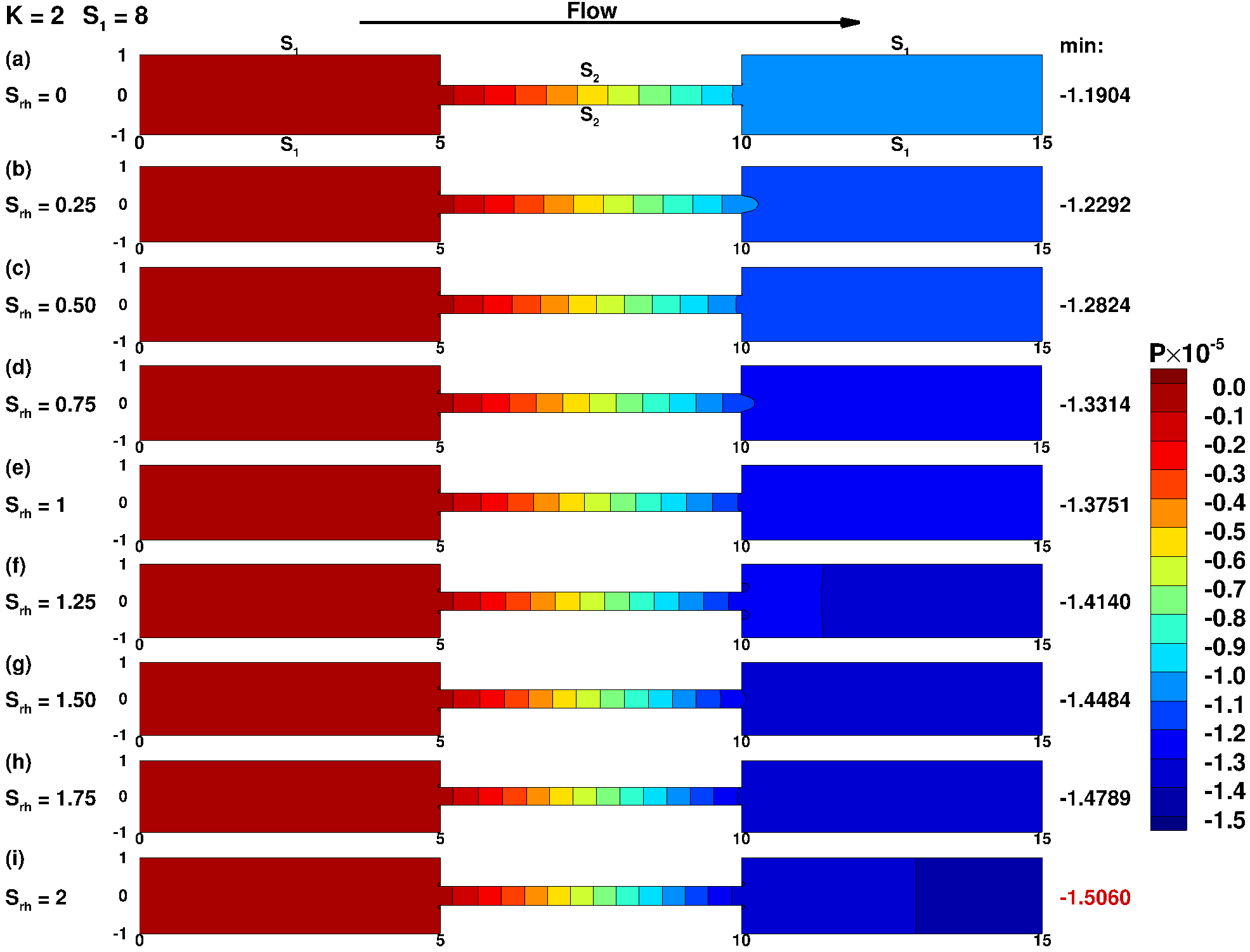}
	\caption{Dimensionless pressure ($P\times10^{-5}$) distribution for $2\le K\le 20$, $4\le S_\text{1}\le 16$, and $0\le S_\text{rh}\le 2$.}
	\label{fig:10}
\end{figure}

\noindent \fig\ref{fig:10} depicts the distribution of pressure ($P$) in the considered microfluidic device for $0\le S_\text{rh}\le 2$, $S_1=8$, and $K=2$. Contour profiles of $P$ are similar for other ranges of dimensionless parameters ($2\le K\le 20$, $4\le S_\text{1}\le 16$, $0\le S_\text{rh}\le 2$) and thus not presented here. The $P$ decreases along the length ($0\le x\le L$) of positively charged channel as expected due to increases both hydrodynamic and electrical resistances with the length of device (\fig\ref{fig:10}). The $P$ decreases with increasing $S_\text{rh}$ (\fig\ref{fig:10}). For instance, minimum value of $P\times10^{-5}$ is recorded as -1.506 (at $S_\text{rh}=2$ for fixed $S_1=8$ and $K=2$) (refer \fig\ref{fig:10}i). However, Overall minimum value of $P\times10^{-5}$ is noted as -1.6343 (at $S_\text{rh}=2$, $S_1=16$, $K=2$). In general, pressure gradient is maximum in the contraction than other regions due reduction in the cross-section area of that section, which increases pressure drop from $\Delta P\propto1/A_\text{c}^2$ (i.e., standard \textit{Hagen-Poiseuille} relation for channel flow), here $A_\text{c}$ is the area of contraction section (\fig\ref{fig:10}). In addition, increment in the pressure gradient in contraction region is noted with increasing charge-heterogeneity $S_\text{rh}$ from 0 to 2 (\fig\ref{fig:10}). It is because increment in $S_2$ (surface charge density of contraction section) is noted with increasing $S_\text{rh}$ from \eqn(\ref{eq:12}), irrespective of $K$ and $S_1$.
\begin{figure}[h]
	\centering\includegraphics[width=1\linewidth]{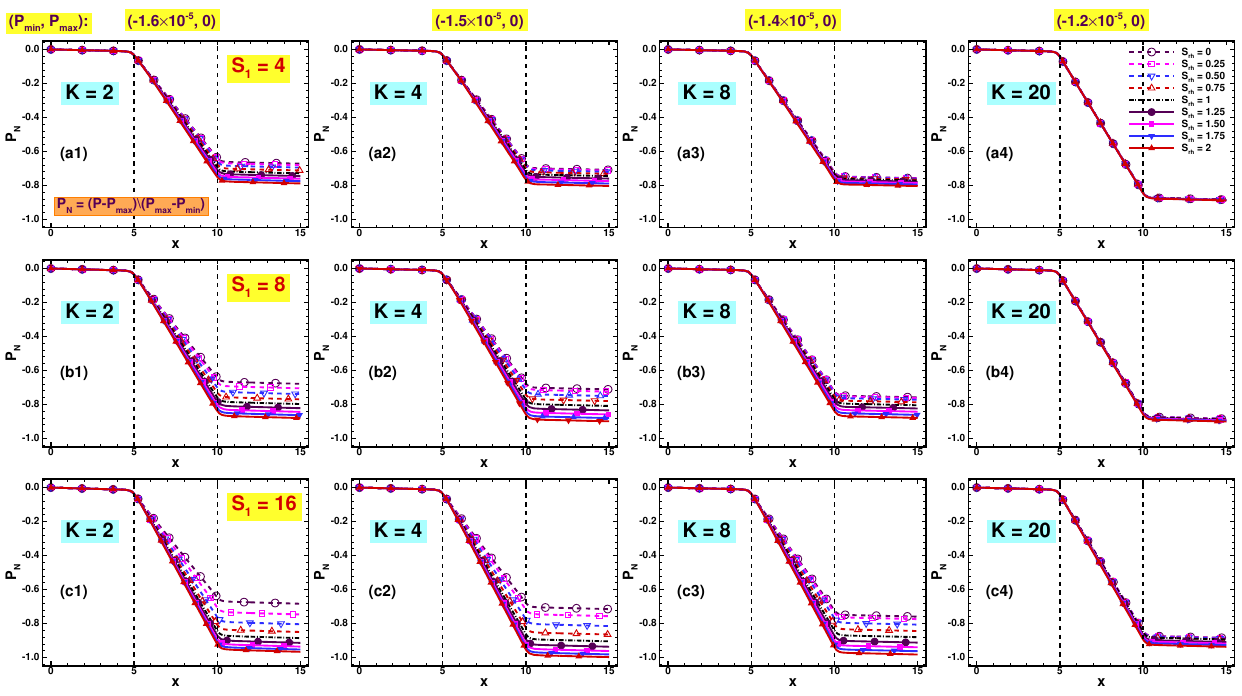}
	\caption{Centreline profiles of normalized pressure ($P_\text{N}$) variation in the heterogeneously charged microfluidic device for $0\le S_\text{rh}\le 2$, $4\le S_\text{1}\le 16$, and $2\le K\le 20$.}
	\label{fig:11a}
\end{figure} 

\noindent Further, extensive analysis of pressure is carried out by normalizing with maximum (\textit{max}) and minimum (\textit{min}) values of $P$ at each $K$; it is defined as $P_\text{N}=(P-P_\text{max})/(P_\text{max}-P_\text{min})$. \fig\ref{fig:11a} shows the centreline ($P_0$ to $P_4$; \fig\ref{fig:1}) profiles of normalized pressure ($P_\text{N}$) in the considered microchannel for $2\le K\le 20$, $4\le S_\text{1}\le 16$, and $0\le S_\text{rh}\le 2$. The $P_\text{N}$ varies in the range of $0\ge P_\text{N}\ge -1$. Centreline profiles of $P_\text{N}$ have shown the similar qualitative behavior as $P$ with the literature \citep{dhakar2022electroviscous,dhakar2023cfd} for $S_\text{rh}=1$. Along the length ($0\le x\le L$) of device, $P_\text{N}$ decreases similar to $P$ in the direction of PDF, irrespective of flow governing parameters ($K$, $S_1$, $S_\text{rh}$) (\fig\ref{fig:11a}). The $P_\text{N}$ decreases with decreasing $K$ due to thickening of the EDL. Further, $|P_\text{N}|$ increases and increasing both $S_1$ and $S_\text{rh}$ (\fig\ref{fig:11a}). In general, contraction has shown maximum gradient of $P_\text{N}$ similar to $P$ than other regions of device (\fig\ref{fig:11a}).

\noindent Subsequently, \tab\ref{tab:1} comprises the pressure drop ($\Delta P$) on the centreline ($P_0$ to $P_4$; \fig\ref{fig:1}) of considered microfluidic device for $2\le K\le 20$, $4\le S_\text{1}\le 16$, and $0\le {S}_\text{rh}\le 2$.  Maximum values of pressure drop ($10^{-5}|\Delta P|$) are underlined data for $0\le S_\text{rh}\le 2$ at each $S_1$ and $K$. The $\Delta P$ variation with $K$ and $S_1$ is same as the literature \citep{davidson2007electroviscous,dhakar2022electroviscous,dhakar2023cfd} for homogeneously charged ($S_\text{rh}=1$) condition. The $\Delta P$ decreases with decreasing $K$ or thickening of EDL; the effect of $K$ on $\Delta P$ is obtained maximum at $S_1=16$ and $S_\text{rh}=2$. For instance, decrement in the values of $\Delta P$ are noted with increasing $K$ from 2 to 20 as (1.36\% \add{(1.0762 to 1.0616)}, 8.94\% \add{(1.1672 to 1.0629)}, 15.50\% \add{(1.2625 to 1.0668)}) and (2.96\% \add{(1.0941 to 1.0617)}, 23.73\% \add{(1.4179 to 1.0815)}, 27.28\% \add{(1.5471 to 1.125)}) for ($S_\text{rh}=0$, 1, 2), respectively at $S_1=4$ and 16 (refer \tab\ref{tab:1}). The maximum variation in $\Delta P$ with $S_1$ is obtained at $K=2$ and $S_\text{rh}=2$. For instance, $\Delta P$ enhances by (1.66\% \add{(1.0762 to 1.0941)}, 21.48\% \add{(1.1672 to 1.4179)}, 22.54\% \add{(1.2625 to 1.5471)}) and (0.01\% \add{(1.0616 to 1.0617)}, 1.75\% \add{(1.0629 to 1.0815)}, 5.46\% \add{(1.0668 to 1.125)}) for ($S_\text{rh}=0$, 1, 2), respectively at $K=2$ and 20 with increasing $S_1$ from 4 to 16 (refer \tab\ref{tab:1}). The change in $\Delta P$ with $S_\text{rh}$ is observed maximum at $K=2$ and $S_1=16$. For instance, $\Delta P$ reduces with decreasing charge-heterogeneity $S_\text{rh}$ from 1 to 0 by (7.80\% \add{(1.1672 to 1.0762)}, 14.91\% \add{(1.2766 to 1.0862)}, 22.84\% \add{(1.4179 to 1.0941)}) and (0.12\% \add{(1.0629 to 1.0616)}, 0.50\% \add{(1.0669 to 1.0616)}, 1.83\% \add{(1.0815 to 1.0617)}) for ($S_1=4$, 8, 16), respectively at $K=2$ and 20; on the other hand, increment in the values of $\Delta P$ are recorded with increasing charge-heterogeneity $S_\text{rh}$ from 1 to 2 as (8.16\% \add{(1.1672 1.2525)}, 10.36\% \add{(1.2766 to 1.4088)}, 9.11\% \add{(1.4179 to 1.5471)}) and (0.37\% \add{(1.0629 to 1.0668)}, 1.35\% \add{(1.0669 to 1.0813)}, 4.02\% \add{(1.0815 1.125)}), respectively at $K=2$ and 20. Overall enhancement in $\Delta P$ is recorded as (17.31\% \add{(1.0762 to 1.2625)}, 29.70\% \add{(1.0862 to 1.4088)}, 41.40\% \add{(1.0941 to 1.5471)}) and (0.49\% \add{(1.0616 to 1.0688)}, 1.86\% \add{(1.0616 to 1.0813)}, 5.96\% \add{(1.0617 to 1.125)}) for ($S_1=4$, 8, 16), respectively at $K=2$ and 20 with overall increasing charge-heterogeneity $S_\text{rh}$ from 0 to 2 ($0\le S_\text{rh}\le 2$) (refer \tab\ref{tab:1}). In summary, increment in $|\Delta P|$ is noted with enhancing both $S_1$ and $S_\text{rh}$. It is because intensified charge attractive forces near the channel walls increase $n^\ast$ (refer section \ref{sec:charge}) and $E_\text{x}$ (i.e., $-\partial U/\partial x$) (refer section \ref{sec:electric}) with increasing $S_1$ and $S_\text{rh}$. Thus, it enhances additional resistance due to electrical force ($\myvec{F_\text{e}}$) on the fluid flow, which increases $|\Delta P|$ from \eqn(\ref{eq:4}) (refer \tab\ref{tab:1}).
\begin{figure}[h]
	\centering\includegraphics[width=1\linewidth]{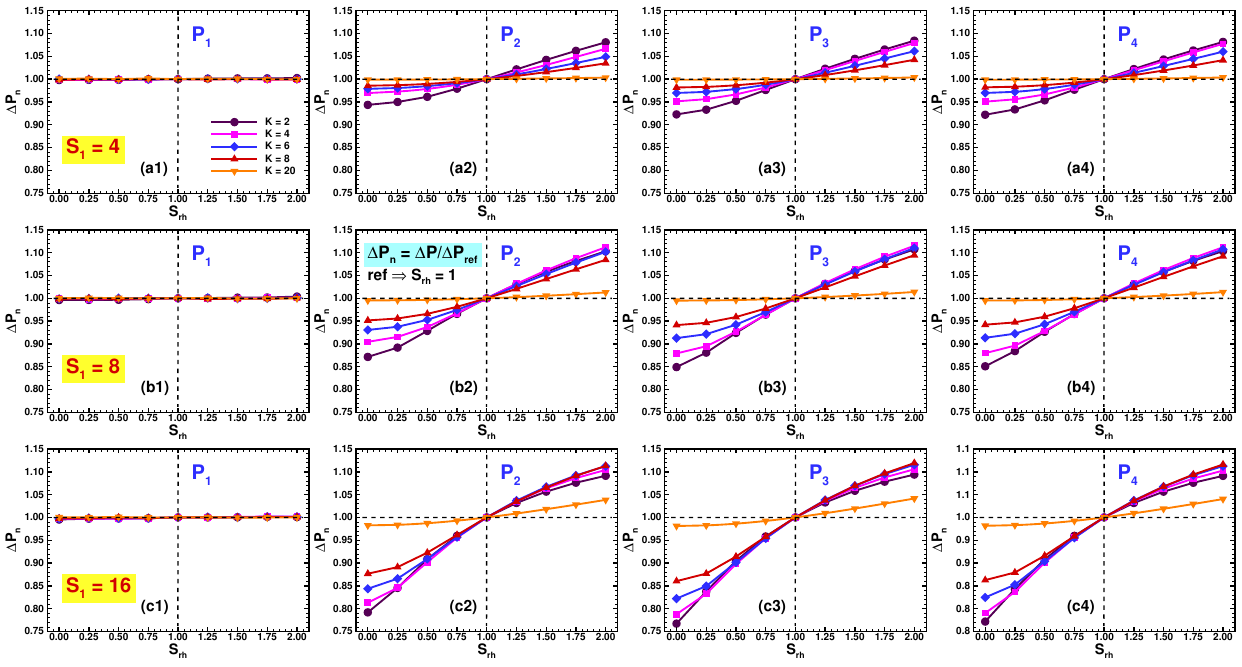}
	\caption{Normalized pressure ($\Delta P_\text{n}$) variation on the centreline locations ($P_1$, $P_2$, $P_3$, $P_4$; \fig\ref{fig:1}) of heterogeneously charged microfluidic device for $2\le K\le 20$, $4\le S_\text{1}\le 16$, and $0\le S_\text{rh}\le 2$.}
	\label{fig:12}
\end{figure} 

\noindent Further, pressure drop in considered geometry is analyzed in detailed by normalizing $\Delta P$ with reference ($S_\text{rh}=1$) condition (refer \eqn\ref{eq:11a}). \fig\ref{fig:12} depicts the normalized pressure drop ($\Delta P_\text{n}$) variation with $S_\text{rh}$ on the centreline locations ($P_1$, $P_2$, $P_3$, $P_4$; \fig\ref{fig:1}) of considered microfluidic device. The $\Delta P_\text{n}$ increases with increasing $K$ for $S_\text{rh}<1$, but it has shown reverse trends for $S_\text{rh}>1$, irrespective of $S_1$ (\fig\ref{fig:12}). The variation in $\Delta P_\text{n}$ with $K$ is maximum at highest $S_1$ and lowest $S_\text{rh}$ (at $P_3$) (\fig\ref{fig:12}). For instance, $\Delta P_\text{n}$ increases maximally by (0.50\% \add{(0.9954 to 1.0003)}, 24.10\% \add{(0.7917 to 0.9825)}, 27.90\% \add{(0.7672 to 0.9812)}, 27.22\% \add{(0.7716 to 0.9817)}) for ($P_1$, $P_2$, $P_3$, $P_4$), respectively with increasing $K$ from 2 to 20 at $S_1=16$ and $S_\text{rh}=0$ (refer \fig\ref{fig:12}). The $\Delta P_\text{n}$ enhances with increasing both $S_1$ and $S_\text{rh}$ or enhancing $S_2$ (\fig\ref{fig:12}). The relative impact of $S_1$ on $\Delta P_\text{n}$ is obtained maximum at lower $K$ and $S_\text{rh}$ (at $P_3$) (\fig\ref{fig:12}). For instance, $\Delta P_\text{n}$ reduces maximally for ($P_1$, $P_2$, $P_3$, $P_4$) by (0.18\% \add{(0.9982 to 0.9964)}, 16.15\% \add{(0.9695 to 0.8129)}, 17.20\% \add{(0.9513 to 0.7878)}, 16.87\% \add{(0.9507 to 0.7903)}), respectively when $S_1$ varies from 4 to 16 at $K=4$ and $S_\text{rh}=0$ (refer \fig\ref{fig:12}). The effect of $S_\text{rh}$ on $\Delta P_\text{n}$ is observed maximum at lowest $K$ and highest $S_1$ (at $P_3$) (\fig\ref{fig:12}). For instance, maximum increment in $\Delta P_\text{n}$ is noted as (0.55\% \add{(0.9954 to 1.0008)}, 37.84\% \add{(0.7917 to 1.0913)}, 42.60\% \add{(0.7672 to 1.0940)}, 41.40\% \add{(0.7716 to 1.0911)}) for ($P_1$, $P_2$, $P_3$, $P_4$), respectively with enhancing $S_\text{rh}$ from 0 to 2 at $K=2$ and $S_1=16$ (refer \fig\ref{fig:12}). In summary, it is noted that maximum change in $\Delta P_\text{n}$ with governing parameters ($K$, $S_1$, $S_\text{rh}$) is obtained at $P_3$ compared to other centreline locations ($P_1$, $P_2$, $P_4$). It attributes that maximum variation in $\Delta U_\text{n}$ at $P_3$ (refer \fig\ref{fig:4} in section \ref{sec:potential}) imposes maximum change in $\Delta P_\text{n}$ (from \eqn\ref{eq:4}) at $P_3$ than other locations ($P_1$, $P_2$, $P_4$) of device (\fig\ref{fig:12}).
\subsection{Electroviscous correction factor ($Y$)}
\label{sec:ECF}
\noindent In electroviscous flows (EVFs), the convective flow of excess charge ($n^\ast$) in the microfluidic device by applied PDF generates an induced electric field strength ($E_\text{x}$) and hence streaming potential. It imposes an additional resistance on the fluid flow in the device and manifests the pressure drop ($\Delta P$) for EVF (i.e., $S_1>0$) that is greater than the pressure drop ($\Delta P_0$) for non-EVF (i.e., $S_1=0$ or $K=\infty$) at fixed volumetric flow rate ($Q$). This relative enhancement in the pressure is measured by the effective or apparent viscosity ($\mu_{\text{eff}}$) and it is quantified as the electroviscous effect (EVE) \citep{davidson2007electroviscous,davidson2008electroviscous,bharti2008steady,bharti2009electroviscous,dhakar2022electroviscous,dhakar2023cfd}. The apparent viscosity ($\mu_{\text{eff}}$) is the viscosity of fluid, needed to obtain pressure drop ($\Delta P$) in absence of electrical forces ($S_1=0$ or $K=\infty$). 

\noindent For low $Re$ steady laminar microfluidic flow, non-linear advection term is negligible in the momentum conservation equation (i.e., \eqn\ref{eq:4}). In turn, relative increment in the pressure drop ($\Delta P/\Delta P_0$) relates to the corresponding relative increment in the viscosity ($\mu_{\text{eff}}/\mu$). Thus, the \textit{electroviscous correction factor} ($Y$) is expressed as follows. 
\begin{gather}
	Y=\frac{\mu_{\text{eff}}}{\mu}=\frac{\Delta P}{\Delta P_{\text{0}}}
	\label{eq:27}
\end{gather}
where $\mu$ is the physical viscosity of liquid.
\begin{figure}[t]
	\centering
	\includegraphics[width=1\linewidth]{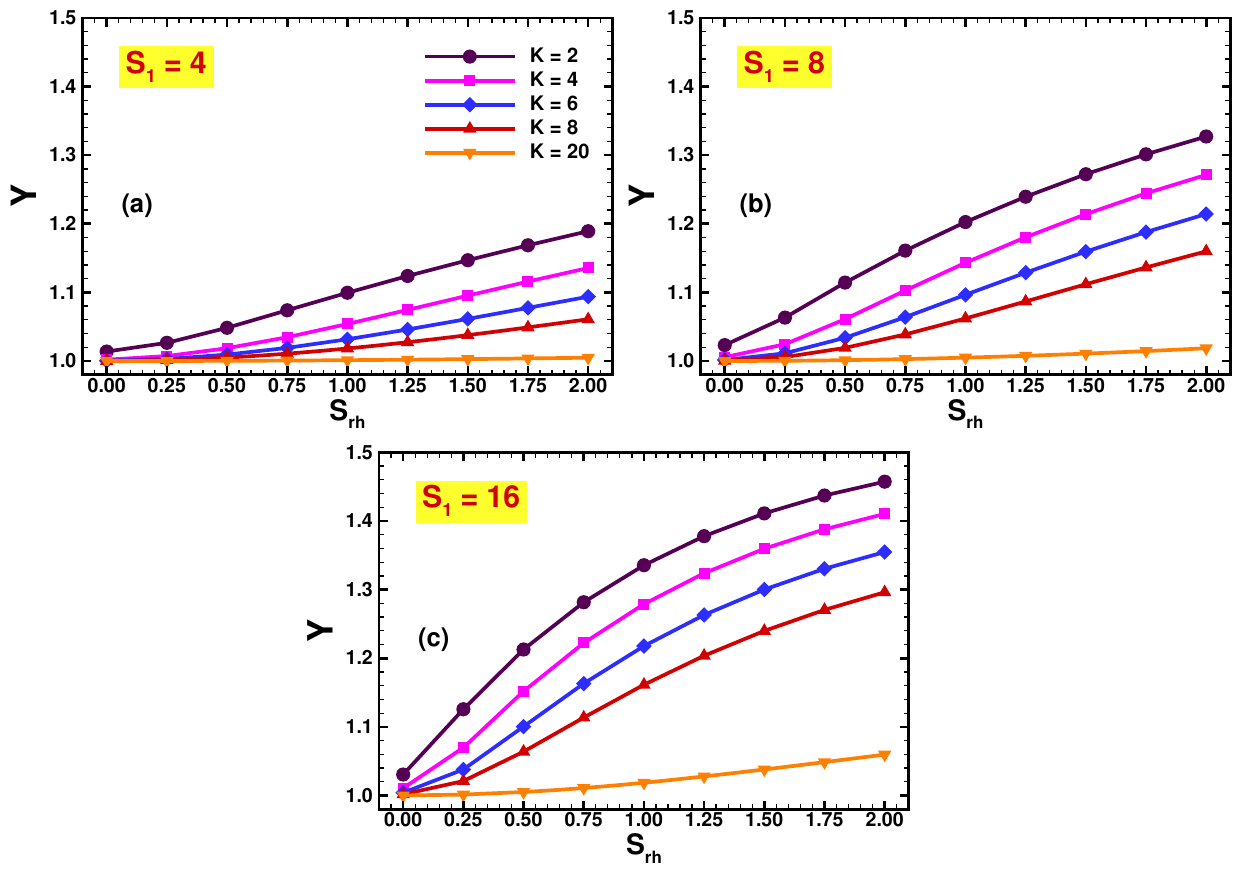}
	\caption{Electroviscous correction factor ($Y$) as a function of $K$, $S_1$ and $S_\text{rh}$.}
	\label{fig:13}
\end{figure} 

\noindent \fig\ref{fig:13} represents the electroviscous correction factor ($Y$) as a function of $K$, $S_1$, and $S_\text{rh}$. In general, electroviscous effects become weak when $Y\rightarrow1$ and absent for $Y=1$. However, electroviscous effects are significantly stronger when $Y>1$. The correction factor has shown complex dependency on $K$, $S_1$, and $S_\text{rh}$. The factor increases with decreasing $K$ or thickening of EDL (\fig\ref{fig:13}). The $Y$ enhances with increasing both $S_1$ and $S_\text{rh}$ (i.e., increasing $S_2$ from \eqn\ref{eq:12}). It is because strengthening in the charge attractive forces near the device walls increases additional resistance on the fluid flow in the channel. Thus, increment in the pressure drop ($\Delta P$) as discussed in section \ref{sec:pressure} is noted, which enhances $Y$ from \eqn(\ref{eq:27}) with increasing both $S_\text{1}$ and $S_\text{rh}$ (\fig\ref{fig:13}). For instance, $Y$ enhances maximally by 24.39\% \add{(1.1158 to 1.3879)} (at $K=4$, $S_\text{rh}=1.75$), 37.52\% \add{(1.0597 to 1.4573)} (at $S_1=16$, $S_\text{rh}=2$), 41.4\% \add{(1.0306 to 1.4573)} (at $K=2$, $S_1=16$) with change in $S_1$ from 4 to 16, $K$ from 20 to 2, $S_\text{rh}$ from 0 to 2, respectively. Further, overall enhancement in $Y$ is recorded as 45.73\% \add{(1 to 1.4573)} (at $K=2$, $S_1=16$, $S_\text{rh}=2$), relative to non-EVF ($S_1=0$ or $K=\infty$) (refer \fig\ref{fig:13}).  

\noindent Electroviscous correction factor ($Y$) functional dependence on dimensionless parameters ($K$, $S_1$, $S_\text{rh}$ and $d_\text{c}=0.25$) is given below. 
\begin{gather}
	Y =  G_1 + (G_2+G_4 K)K + (G_3 +G_5 S_\text{rh})S_\text{rh} + G_6 KS_\text{rh}
	\label{eq:28}
	\\\text{where}\qquad 
	G_{\text{i}} = \sum_{{j}=1}^3 M_{\text{ij}} \gamma^{({j}-1)}\quad\text{and}\quad \gamma = S_\text{1}^{-1}, \quad 1\le i\le 6 \nonumber
\end{gather}
In \eqn\ref{eq:28} correlation coefficients ($M_{\text{ij}}$) are statistically obtained, for 135 data points, as 
\begin{gather*}
	M = \begin{bmatrix}
	1.1140	&	-0.3407	&	0.3567	\\
	-0.0198	&	-0.0317	&	0.2145	\\
	0.5483	&	-4.2297	&	9.2063	\\
	0.0005	&	0.0050	&	-0.0181	\\
	-0.1041	&	1.2211	&	-3.0704	\\
	-0.0131	&	0.0466	&	-0.0489		
	\end{bmatrix} 
\end{gather*}
by performing the non-linear regression analysis using DataFit (trial version) with ($\delta_{\text{min}}$, $\delta_{\text{max}}$, $\delta_{\text{avg}}$, $R^2$) as (-3.29\%, 2.54\%, 0.06\%, 99.03\%) for given ranges of conditions.
\subsection{Pseudo-analytical model}
\noindent The hydrodynamic characteristics such as pressure have shown complex dependency on the flow governing parameters as discussed in the preceding discussion, which calculated numerically (from \eqn\ref{eq:4}) in the microfluidic device. However, it can be predict analytically using a simpler pseudo-analytical model for wide ranges of flow governing parameters and easy uses for designing the relevant microfluidic devices for practical applications. Earlier studies \citep{davidson2007electroviscous,bharti2008steady,dhakar2022electroviscous,dhakar2023cfd} have discussed and presented the pseudo-analytical model to calculate the pressure drop ($\Delta P$) in liquid flow through symmetrically/asymmetrically charged contraction-expansion ($d_\text{c}=0.25$) slit/cylinder microfluidic devices. These studies \citep{davidson2007electroviscous,bharti2008steady,dhakar2022electroviscous,dhakar2023cfd} have predicted the pressure drop in steady laminar fully-developed flow of Newtonian and incompressible electrolyte liquid through contraction-expansion microfluidic device by summation of the pressure drop in independently uniform (i.e., $\Delta P_\text{u}$, $\Delta P_\text{c}$, and $\Delta P_\text{d}$) rectangular section by standard \textit{Hagen-Poiseuille equation}, and excess pressure drop (i.e., $\Delta P_\text{e}$) due to sudden contraction/expansion calculated by pressure drop through thin ($d_\text{c}<<1$) orifice \citep{Sisavath2002,davidson2007electroviscous,Pimenta2020}. It is expressed as follows.
\begin{gather}
	\Delta P_{\text{0,m}}=\left(\sum_{i=u,c,d}\Delta P_{\text{0,i}}\right) +\Delta P_{\text{0,e}}
	\label{eq:34}
\end{gather}
\begin{gather}
	\Delta P_{\text{0,i}} = \left(\frac{3}{Re}\right)\frac{\Delta L_{\text{i}}}{d_\text{i}^3};\qquad 
	\Delta P_{\text{0,e}} =\frac{16}{\pi d_{\text{c}}^2Re};\qquad \text{where} \qquad  d_\text{i}=\frac{W_\text{i}}{W} 
	\label{eq:35}
\end{gather}
where subscripts '0', $u$, $d$, and $c$ denote the non-EVF condition ($S_1=0$ or $K=\infty$), upstream, downstream, and contraction sections, respectively. The $Re$ is defined in \eqn\ref{eq:1} and $d_{\text{c}}$ is the contraction ratio.

\noindent Further, above model (\eqns\ref{eq:34}) is extended, and generalized simpler pseudo-analytical model (\eqn\ref{eq:38}) is proposed to estimate the pressure drop in the electroviscous flow ($S_1>0$) through heterogeneously charged ($S_\text{rh}\neq1$) contraction-expansion ($d_\text{c}=0.25$) slit microfluidic device and it is expressed as follows.
\begin{gather}
\Delta P_{\text{m}}= \Gamma_\text{rh} \Delta P_{0,\text{m}}
=
\left(\frac{3\Gamma_\text{rh}}{Re}\right)\left(L_{\text{u}} +  \frac{L_{\text{c}}}{d_{\text{c}}^3} + L_{\text{d}} + \frac{16}{3\pi d_{\text{c}}^2} \right)
\label{eq:38}
\end{gather}
The correction coefficient ($\Gamma_\text{rh}$, \eqn\ref{eq:38}) accounts for influence of electroviscous ($S_1>0$) and charge-heterogeneity ($S_\text{rh}\neq1$) effects on the pressure drop ($\Delta P_{0,\text{m}}$) as follows.
\begin{gather}
	\Gamma_\text{rh} = G_1 + (G_2+G_4 K)K + (G_3+G_5 S_\text{rh})S_\text{rh} + G_6 KS_\text{rh}
	\\\text{where}\qquad 
	G_{\text{i}} = \sum_{{j}=1}^3 M_{\text{ij}} \gamma^{({j}-1)}\quad\text{and}\quad \gamma = S_\text{1}^{-1}, \quad 1\le i\le 6 \nonumber
\end{gather}
The correlation coefficients ($M_\text{ij}$) are statistically obtained, for 135 data points, as 
\begin{gather*}
	M = \begin{bmatrix}
	1.1037	&	-0.338	&	0.3546	\\
	-0.0196	&	-0.032	&	0.2126	\\
	0.5431	&	-4.189	&	9.1155	\\
	4.9\times10^{-4} &	0.005 &	-0.0181	\\
	-0.1030	&	1.208	&	-3.0389	\\
	-0.0130	&	0.046	&	-0.0484		
	\end{bmatrix} 
\end{gather*}
with ($\delta_{\text{min}}$, $\delta_{\text{max}}$, $\delta_{\text{avg}}$, $R^2$) as (-3.31\%, 2.54\%, -0.00\%, 99.03\%) for given ranges of conditions.

\noindent Subsequently, \eqns(\ref{eq:34}) and (\ref{eq:38}) are further extended to calculate the electroviscous correction factor and it is expressed as follows.
\begin{gather}
	Y = \frac{\Delta P_\text{m}}{\Delta P_\text{0,m}}
	\label{eq:39}
\end{gather}
\begin{figure}[htbp]
	\centering
	\subfigure[pressure drop] {\includegraphics[width=0.49\linewidth]{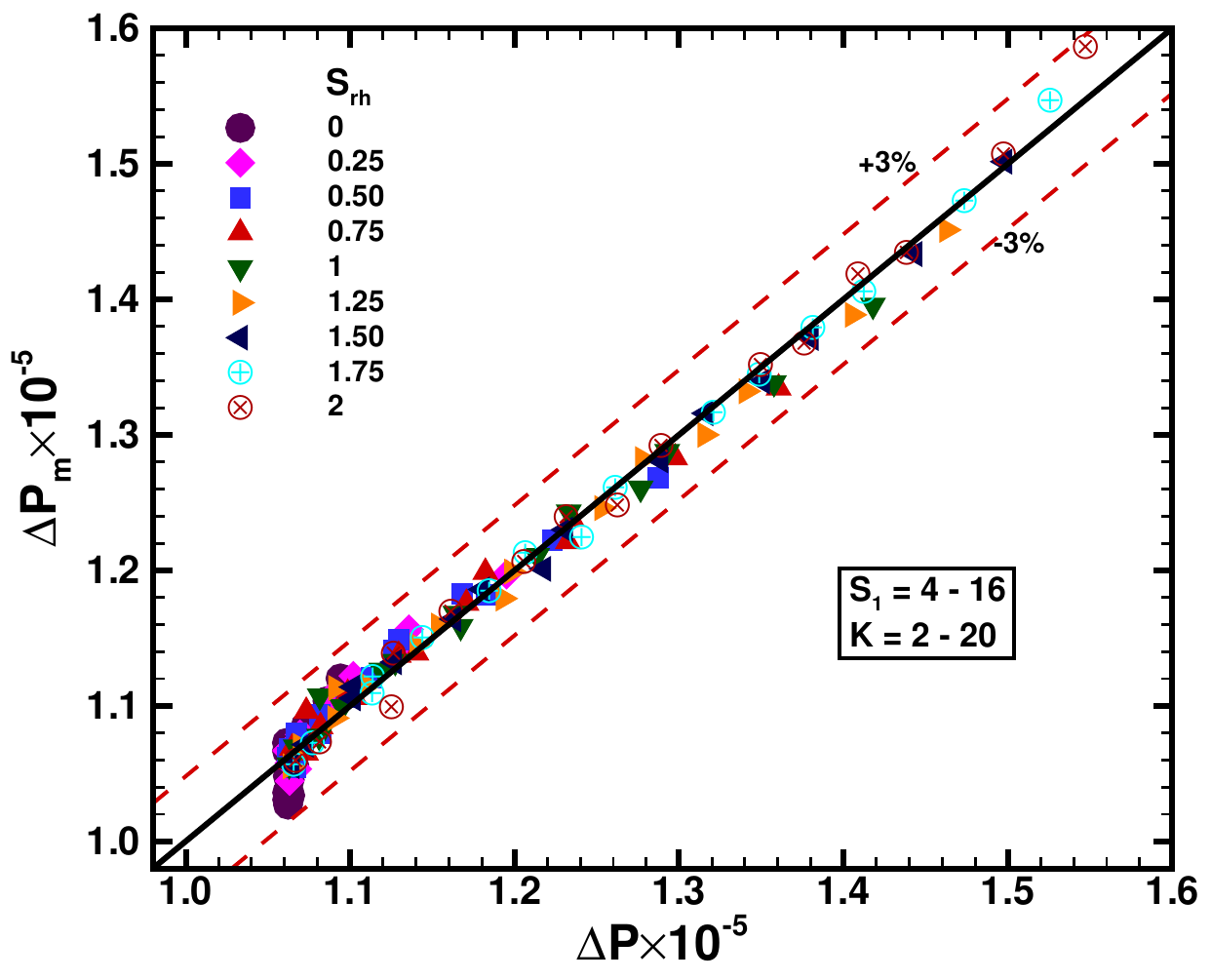}}
	\subfigure[electroviscous correction factor]
	{\includegraphics[width=0.49\linewidth]{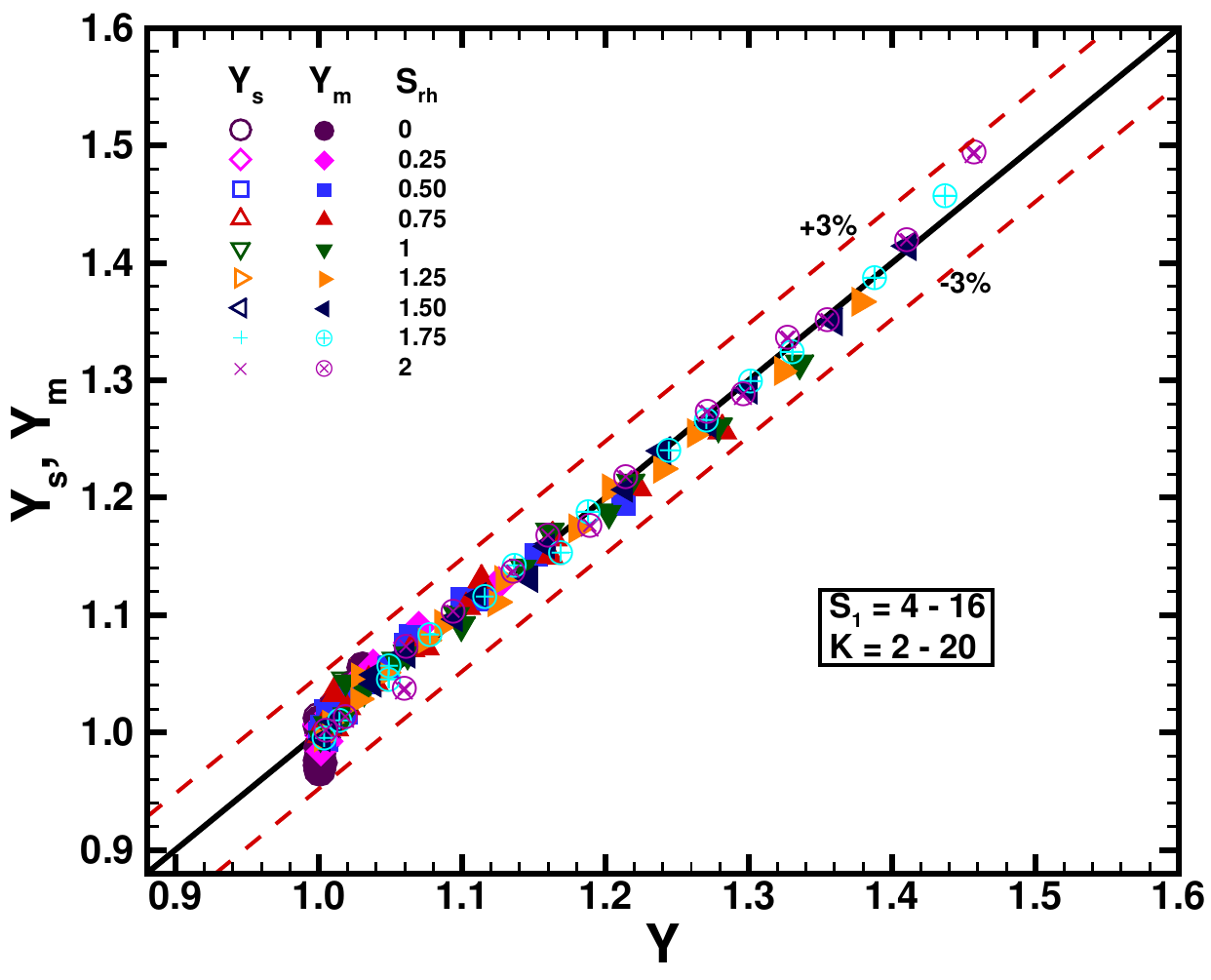}}
	\caption{Parity chart between the numerical and pseudo-analytical values of the (a) pressure drop, $\Delta P$ vs. $\Delta P_{\text{m}}$, (b) electroviscous correction factor, $Y$ vs. $Y_{\text{s}}$ and $Y_{\text{m}}$, for the considered parameters ($K$, $S_1$, $S_\text{rh}$).}
	\label{fig:14}
\end{figure} 
\noindent \fig\ref{fig:14}(a) and (b) represent the parity charts for pressure drop ($\Delta P_\text{m}$ vs $\Delta P$) and electroviscous correction factor ($Y_\text{m}$ vs $Y$) calculated from simpler pseudo-analytical model (\eqns\ref{eq:38} and \ref{eq:39}) and numerical simulations, respectively. This analytical model overestimates the pressure drop, $\Delta P$ and correction factor, $Y$ by $\pm$3\% compared to present numerical results. Further, difference between the predicted model and numerical values is reduced with decreasing both $S_1$ and $S_\text{rh}$ and increasing $K$ or thinning of the EDL.
\section{Concluding remarks}
\noindent In the present work, charge-heterogeneity (CH) effects on electroviscous (EV) flow of electrolyte liquid through contraction-expansion slit microfluidic device have investigated using numerical simulations. Mathematical model consisting of the Poisson's, Nernst-Planck, Navier-Stokes, and current continuity equations is solved numerically by using the finite element method (FEM). Numerical results are presented for total electrical potential, excess charge, induced electric field strength, pressure, and electroviscous correction factor for wide ranges of conditions ($Re=10^{-2}$, $\mathit{Sc}=10^{-3}$, $\beta=2.34\times10^{-4}$, $2\le K\le 20$, $4\le S_1\le 16$, $0\le S_\text{rh}\le 2$).

\noindent  Charge-heterogeneity (CH) significantly affects the hydrodynamic characteristics in the microfluidic device. Maximum enhancement in $|\Delta U|$ and $|\Delta P|$ are recorded as 3511.45\% \add{(0.2127 to 7.6801)} (at $K=20$, $S_1=4$) and 41.4\% \add{(1.0941 to 1.5471)} (at $K=2$, $S_1=16$), respectively with overall increasing charge-heterogeneity $S_\text{rh}$ from 0 to 2. The $Y$ (i.e., $\Delta P/\Delta P_0=\mu_{\text{eff}}/\mu$) increases maximally by 24.39\% \add{(1.1158 to 1.3879)} (at $K=4$, $S_\text{rh}=1.75$), 37.52\% \add{(1.0597 to 1.4573)} (at $S_1=16$, $S_\text{rh}=2$), 41.4\% \add{(1.0306 to 1.4573)} (at $K=2$, $S_1=16$) with the variation in $S_1$ from 4 to 16, $K$ from 20 to 2, $S_\text{rh}$ from 0 to 2, respectively. Further, overall increment in $Y$ is noted as 45.73\% \add{(1 to 1.4573)} (at $K=2$, $S_1=16$, $S_\text{rh}=2$), relative to non-EVF ($S_1=0$ or $K=\infty$). Thus, charge-heterogeneity enhances electroviscous effects in the microfluidic devices.

\noindent A simple predictive pseudo-analytical model is developed to calculate the pressure drop, $\Delta P$ (and electroviscous correction factor, $Y$) in heterogenously charged contraction-expansion microfluidic device. It estimates both pressure drop and correction factor within $\pm$3\% compared to present numerical results. Thus, mathematical correlations and robustness of this simple pseudo-analytical model for wide ranges of flow governing parameters ($K$, $S_1$, $S_\text{rh}$), enable the present numerical results uses for developing the reliable, efficient, and precisely controlled microfluidic devices for practical applications.
\section*{Declaration of Competing Interest}
\noindent 
The authors declare that they have no known competing financial interests or personal relationships that could have appeared to influence the work reported in this paper.
%
\section*{Acknowledgements}
\noindent 
The authors acknowledge the infrastructural, computing resources, and software license support from the Indian Institute of Technology Roorkee. 
JD is thankful to the Department of Higher Education, Ministry of Education (MoE), Government of India (GoI) for the providence of research fellowship. 
%
\begin{spacing}{1.5}
\fontsize{10}{10pt}\selectfont
 \nomenclature[g0]{\textit{Greek letters}}{}
 \nomenclature[d0]{\textit{Dimensionless groups}}{}
 \nomenclature[s0]{\textit{Subscripts and Superscripts}}{}
 \nomenclature[z0]{\textit{Abbreviations}}{}
%
\nomenclature[zcfd]{CFD}{computational fluid dynamics}
\nomenclature[zedl]{EDL}{electrical double layer}
\nomenclature[zevf]{EVF}{electroviscous flow}
\nomenclature[zfem]{FEM}{finite element method}
\nomenclature[zfvm]{FVM}{finite volume method}
\nomenclature[zpdes]{PDEs}{partial differential equations}
\nomenclature[zpdet]{PDF}{pressure-driven flow}
\nomenclature[zsaes]{SAEs}{simultaneous algebraic equations}
%
\nomenclature[aD]{$\mathcal{D}$}{diffusivity of the positive and negative ions, assumed equal ($\mathcal{D}_{+}=\mathcal{D}_{-}=\mathcal{D}$), m$^2$/s}
\nomenclature[adc]{$d_{\text{c}}$}{contraction ratio ($=W_{\text{c}}/W$), --}
\nomenclature[aDj]{$\mathcal{D}_{j}$}{diffusivity of the ions of type j, m$^2$/s}
\nomenclature[ae]{$e$}{elementary charge of a proton ($=1.602176634\times 10^{-19}$), C or A.s}
\nomenclature[aE]{$E_{\text{x}}$}{induced electric field strength, V/m or --}
\nomenclature[afj]{$\mathbf{f_\text{j}}$}{flux density of the ions of type j (\eqn\ref{eq:6}), 1/(m$^2$.s)}
\nomenclature[aIc]{$I_{\text{c}}$}{conduction current density (\eqn\ref{eq:7}), A/m$^2$ or --}
\nomenclature[aId]{$I_{\text{d}}$}{diffusion current density (\eqn\ref{eq:7}), A/m$^2$ or --}
\nomenclature[aIs]{$I_{\text{s}}$}{streaming current density (\eqn\ref{eq:7}), A/m$^2$ or --}
\nomenclature[akB]{$k_{\text{B}}$}{Boltzmann constant ($=1.380649\times 10^{-23}$), J/K}
\nomenclature[aLc]{$L_{\text{c}}$}{length of contraction section, m or --}
\nomenclature[aLd]{$L_{\text{d}}$}{length of downstream outlet section, m or --}
\nomenclature[aLu]{$L_{\text{u}}$}{length of upstream inlet section, m or --}
\nomenclature[an+]{$n_{+}$}{local number density of positive ions (\eqn\ref{eq:6}), 1/m$^3$ or --}
\nomenclature[an-]{$n_{-}$}{local number density of negative ions (\eqn\ref{eq:6}), 1/m$^3$ or --}
\nomenclature[an0]{$n_{0}$}{bulk density of the ions of type j, 1/m$^3$}
\nomenclature[anj]{$n_{j}$}{local number density of the ions of type j, 1/m$^3$}
\nomenclature[ans]{$n^*$}{excess charge ($=n_{+}-n_{-}$), 1/m$^3$ or --}
\nomenclature[aP]{$P$}{pressure, Pa or --}
\nomenclature[aT]{$T$}{temperature, K}
\nomenclature[aU]{$U$}{total electrical potential, V or --}
\nomenclature[aV]{$\mathbf{V}$}{velocity vector, m/s or --}
\nomenclature[aVa]{$\overline{V}$}{average velocity of the fluid at the inlet, m/s}
\nomenclature[aVx]{$V_x$}{x-component of the velocity, m/s or --}
\nomenclature[aVy]{$V_y$}{y-component of the velocity, m/s or --}
\nomenclature[aW]{$W$}{cross-sectional width of inlet and outlet sections, m}
\nomenclature[aWc]{$W_{\text{c}}$}{cross-sectional width of contraction section, m}
\nomenclature[ax]{$x$}{streamwise coordinate, --}
\nomenclature[ay]{$y$}{transverse coordinate, --}
\nomenclature[aY]{$Y$}{electroviscous correction factor (\eqns\ref{eq:27}, and \ref{eq:39}), --}
\nomenclature[azj]{$z_{j}$}{valency of the ions of type j, assumed equal ($z_{+}=z_{-}=z$), --}
%
%
\nomenclature[gdP]{$\Delta P$}{pressure drop (\eqns\ref{eq:38}), --}
\nomenclature[geps0]{$\varepsilon_{\text{0}}$}{permittivity of free space (i.e. vaccum), F/m or C/(V.m)}
\nomenclature[gepsr]{$\varepsilon_{\text{r}}$}{dielectric constant (or absolute permittivity or relative permittivity) of the electrolyte liquid, --}
\nomenclature[glambdad]{$\lambda_{\text{D}}$}{Debye length $\left(=\sqrt{\frac{\varepsilon_{\text{0}}\varepsilon_{\text{r}} k_{\text{b}}T}{z^2e^2n_{\text{0}}}}\right)$, m}
\nomenclature[gmu]{$\mu$}{viscosity, Pa.s}
\nomenclature[gmueff]{$\mu_\text{eff}$}{effective or apparent viscosity, Pa.s}
\nomenclature[gpsi]{$\psi$}{EDL potential, V or --}
\nomenclature[grho]{$\rho$}{density of fluid, kg/m$^3$}
\nomenclature[grhoe]{$\rho_{\text{e}}$}{charge density of liquid, C/m$^3$}
\nomenclature[gsigmab]{$\sigma_\text{2}$}{contraction section walls surface charge density, C/m$^2$}
\nomenclature[gsigma]{$\sigma_\text{1}$}{upstream/downstream sections walls surface charge density, C/m$^2$}
%
%
\nomenclature[dbeta]{$\mathit{\beta}$}{liquid parameter (\eqn\ref{eq:1}), --}
\nomenclature[dK]{$\mathit{K}$}{inverse Debye length (\eqn\ref{eq:1}), --}
\nomenclature[dPe]{$Pe$}{Peclet number ($={Re}~\mathit{Sc}$) (\eqn\ref{eq:1}), --}
\nomenclature[dRe]{$Re$}{Reynolds number (\eqn\ref{eq:1}), --}
\nomenclature[dSd]{$\mathit{S_2}$}{contraction section walls surface charge density (\eqn\ref{eq:11}), --}
\nomenclature[dSc]{$\mathit{Sc}$}{Schmidt number (\eqn\ref{eq:1}), --}
\nomenclature[dSb]{$\mathit{S_\text{rh}}$}{charge-heterogeneity ratio (\eqn\ref{eq:12}), --}
\nomenclature[dSa]{$\mathit{S_1}$}{upstream/downstream sections walls surface charge density (\eqn\ref{eq:11}), --}
%
\nomenclature[sz]{$0$}{without electroviscous effects}
\nomenclature[sc]{$c$}{contraction}
\nomenclature[sd]{$d$}{downstream}
\nomenclature[se]{$e$}{extra or excess}
\nomenclature[sm]{$m$}{mathematical}
\nomenclature[ss]{$s$}{statistical}
\nomenclature[su]{$u$}{upstream}
%

\printnomenclature
\end{spacing}
\bibliography{references}

\begin{thebibliography}{64}
\providecommand{\natexlab}[1]{#1}
\providecommand{\url}[1]{\texttt{#1}}
\providecommand{\urlprefix}{URL }
\expandafter\ifx\csname urlstyle\endcsname\relax
  \providecommand{\doi}[1]{doi:\discretionary{}{}{}#1}\else
  \providecommand{\doi}[1]{doi:\discretionary{}{}{}\begingroup
  \urlstyle{rm}\url{#1}\endgroup}\fi
\providecommand{\bibinfo}[2]{#2}

\bibitem[{Bhushan(2010)}]{bhushan2007springer}
\bibinfo{editor}{B.~Bhushan} (Ed.), {\color{red} {\color{red}
  \bibinfo{title}{Handbook of Nanotechnology}}}, \bibinfo{publisher}{Springer,
  Berlin, Heidelberg}, \bibinfo{edition}{3} edn., \bibinfo{year}{2010}.

\bibitem[{Li(2008)}]{li2008encyclopedia}
\bibinfo{editor}{D.~Li} (Ed.), {\color{red} {\color{red}
  \bibinfo{title}{Encyclopedia of Microfluidics and Nanofluidics}}},
  \bibinfo{publisher}{Springer, Boston, MA}, \bibinfo{year}{2008}.

\bibitem[{Lin(2011)}]{lin2011microfluidics}
\bibinfo{editor}{B.~Lin} (Ed.), {\color{red} {\color{red}
  \bibinfo{title}{Microfluidics: Technologies and Applications}}},
  \bibinfo{publisher}{Springer, Berlin, Heidelberg}, \bibinfo{edition}{1} edn.,
  \bibinfo{year}{2011}.

\bibitem[{Nguyen et~al.(2013)Nguyen, Shaegh, Kashaninejad, and
  Phan}]{nguyen2013design}
\bibinfo{author}{N.-T. Nguyen}, \bibinfo{author}{S.~A.~M. Shaegh},
  \bibinfo{author}{N.~Kashaninejad}, \bibinfo{author}{D.-T. Phan},
  \bibinfo{title}{Design, fabrication and characterization of drug delivery
  systems based on lab-on-a-chip technology}, {\color{red}
  \bibinfo{journal}{Advanced Drug Delivery Reviews}}
  \bibinfo{volume}{65}~(\bibinfo{number}{11-12}) (\bibinfo{year}{2013})
  \bibinfo{pages}{1403--1419}.

\bibitem[{Bruijns et~al.(2016)Bruijns, Van~Asten, Tiggelaar, and
  Gardeniers}]{bruijns2016microfluidic}
\bibinfo{author}{B.~Bruijns}, \bibinfo{author}{A.~Van~Asten},
  \bibinfo{author}{R.~Tiggelaar}, \bibinfo{author}{H.~Gardeniers},
  \bibinfo{title}{Microfluidic devices for forensic DNA analysis: A review},
  {\color{red} \bibinfo{journal}{Biosensors}}
  \bibinfo{volume}{6}~(\bibinfo{number}{3}) (\bibinfo{year}{2016})
  \bibinfo{pages}{41}.

\bibitem[{Li and Zhou(2021)}]{li2021microfluidic}
\bibinfo{author}{X.~J. Li}, \bibinfo{author}{Y.~Zhou}, {\color{red}
  {\color{red} \bibinfo{title}{Microfluidic Devices for Biomedical
  Applications}}}, \bibinfo{publisher}{Woodhead Publishing},
  \bibinfo{year}{2021}.

\bibitem[{Hunter(1981)}]{hunter2013zeta}
\bibinfo{author}{R.~J. Hunter}, {\color{red} {\color{red} \bibinfo{title}{Zeta
  Potential in Colloid Science: Principles and Applications}}},
  \bibinfo{publisher}{Academic Press}, \bibinfo{year}{1981}.

\bibitem[{Li(2001)}]{li2001electro}
\bibinfo{author}{D.~Li}, \bibinfo{title}{Electro-viscous effects on
  pressure-driven liquid flow in microchannels}, {\color{red}
  \bibinfo{journal}{Colloids and Surfaces A: Physicochemical and Engineering
  Aspects}} \bibinfo{volume}{195}~(\bibinfo{number}{1-3})
  (\bibinfo{year}{2001}) \bibinfo{pages}{35--57}.

\bibitem[{Davidson and Harvie(2007)}]{davidson2007electroviscous}
\bibinfo{author}{M.~R. Davidson}, \bibinfo{author}{D.~J.~E. Harvie},
  \bibinfo{title}{Electroviscous effects in low Reynolds number liquid flow
  through a slit-like microfluidic contraction}, {\color{red}
  \bibinfo{journal}{Chemical Engineering Science}}
  \bibinfo{volume}{62}~(\bibinfo{number}{16}) (\bibinfo{year}{2007})
  \bibinfo{pages}{4229--4240}.

\bibitem[{Dhakar and Bharti(2022{\natexlab{a}})}]{dhakar2022electroviscous}
\bibinfo{author}{J.~Dhakar}, \bibinfo{author}{R.~P. Bharti},
  \bibinfo{title}{Electroviscous effects in charge-dependent slip flow of
  liquid electrolytes through a charged microfluidic device}, {\color{red}
  \bibinfo{journal}{Chemical Engineering and Processing-Process
  Intensification}}  (\bibinfo{year}{2022}{\natexlab{a}})
  \bibinfo{pages}{109041}.

\bibitem[{Dhakar and Bharti(2023{\natexlab{a}})}]{dhakar2023cfd}
\bibinfo{author}{J.~Dhakar}, \bibinfo{author}{R.~P. Bharti},
  \bibinfo{title}{Electroviscous effects in pressure-driven flow of electrolyte
  liquid through an asymmetrically charged non-uniform microfluidic device},
  {\color{red} \bibinfo{journal}{Journal of the Taiwan Institute of Chemical
  Engineers}} \bibinfo{volume}{153} (\bibinfo{year}{2023}{\natexlab{a}})
  \bibinfo{pages}{105230}.

\bibitem[{Atten and Honda(1982)}]{atten1982electroviscous}
\bibinfo{author}{P.~Atten}, \bibinfo{author}{T.~Honda}, \bibinfo{title}{The
  electroviscous effect and its explanation I—The electrohydrodynamic origin;
  study under unipolar DC injection}, {\color{red} \bibinfo{journal}{Journal of
  Electrostatics}} \bibinfo{volume}{11}~(\bibinfo{number}{3})
  (\bibinfo{year}{1982}) \bibinfo{pages}{225--245}.

\bibitem[{Rice and Whitehead(1965)}]{rice1965electrokinetic}
\bibinfo{author}{C.~L. Rice}, \bibinfo{author}{R.~Whitehead},
  \bibinfo{title}{Electrokinetic flow in a narrow cylindrical capillary},
  {\color{red} \bibinfo{journal}{Journal of Physical Chemistry}}
  \bibinfo{volume}{69}~(\bibinfo{number}{11}) (\bibinfo{year}{1965})
  \bibinfo{pages}{4017--4024}.

\bibitem[{Levine et~al.(1975)Levine, Marriott, Neale, and
  Epstein}]{levine1975theory}
\bibinfo{author}{S.~Levine}, \bibinfo{author}{J.~Marriott},
  \bibinfo{author}{G.~Neale}, \bibinfo{author}{N.~Epstein},
  \bibinfo{title}{Theory of electrokinetic flow in fine cylindrical capillaries
  at high zeta-potentials}, {\color{red} \bibinfo{journal}{Journal of Colloid
  and Interface Science}} \bibinfo{volume}{52}~(\bibinfo{number}{1})
  (\bibinfo{year}{1975}) \bibinfo{pages}{136--149}.

\bibitem[{Bowen and Jenner(1995)}]{bowen1995electroviscous}
\bibinfo{author}{W.~R. Bowen}, \bibinfo{author}{F.~Jenner},
  \bibinfo{title}{Electroviscous effects in charged capillaries}, {\color{red}
  \bibinfo{journal}{Journal of Colloid and Interface Science}}
  \bibinfo{volume}{173}~(\bibinfo{number}{2}) (\bibinfo{year}{1995})
  \bibinfo{pages}{388--395}.

\bibitem[{Brutin and Tadrist(2005)}]{brutin2005modeling}
\bibinfo{author}{D.~Brutin}, \bibinfo{author}{L.~Tadrist},
  \bibinfo{title}{Modeling of surface-fluid electrokinetic coupling on the
  laminar flow friction factor in microtubes}, {\color{red}
  \bibinfo{journal}{Microscale Thermophysical Engineering}}
  \bibinfo{volume}{9}~(\bibinfo{number}{1}) (\bibinfo{year}{2005})
  \bibinfo{pages}{33--48}.

\bibitem[{Bharti et~al.(2009)Bharti, Harvie, and
  Davidson}]{bharti2009electroviscous}
\bibinfo{author}{R.~P. Bharti}, \bibinfo{author}{D.~J.~E. Harvie},
  \bibinfo{author}{M.~R. Davidson}, \bibinfo{title}{Electroviscous effects in
  steady fully developed flow of a power-law liquid through a cylindrical
  microchannel}, {\color{red} \bibinfo{journal}{International Journal of Heat
  and Fluid Flow}} \bibinfo{volume}{30}~(\bibinfo{number}{4})
  (\bibinfo{year}{2009}) \bibinfo{pages}{804--811}.

\bibitem[{Jing and Pan(2016)}]{jing2016electroviscous}
\bibinfo{author}{D.~Jing}, \bibinfo{author}{Y.~Pan},
  \bibinfo{title}{Electroviscous effect and convective heat transfer of
  pressure-driven flow through microtubes with surface charge-dependent slip},
  {\color{red} \bibinfo{journal}{International Journal of Heat and Mass
  Transfer}} \bibinfo{volume}{101} (\bibinfo{year}{2016})
  \bibinfo{pages}{648--655}.

\bibitem[{Burgreen and Nakache(1964)}]{burgreen1964electrokinetic}
\bibinfo{author}{D.~Burgreen}, \bibinfo{author}{F.~Nakache},
  \bibinfo{title}{Electrokinetic flow in ultrafine capillary slits},
  {\color{red} \bibinfo{journal}{Journal of Physical Chemistry}}
  \bibinfo{volume}{68}~(\bibinfo{number}{5}) (\bibinfo{year}{1964})
  \bibinfo{pages}{1084--1091}.

\bibitem[{Mala et~al.(1997{\natexlab{a}})Mala, Li, Werner, Jacobasch, and
  Ning}]{mala1997flow}
\bibinfo{author}{G.~M. Mala}, \bibinfo{author}{D.~Li},
  \bibinfo{author}{C.~Werner}, \bibinfo{author}{H.-J. Jacobasch},
  \bibinfo{author}{Y.~B. Ning}, \bibinfo{title}{Flow characteristics of water
  through a microchannel between two parallel plates with electrokinetic
  effects}, {\color{red} \bibinfo{journal}{International Journal of Heat and
  Fluid Flow}} \bibinfo{volume}{18}~(\bibinfo{number}{5})
  (\bibinfo{year}{1997}{\natexlab{a}}) \bibinfo{pages}{489--496}.

\bibitem[{Mala et~al.(1997{\natexlab{b}})Mala, Li, and Dale}]{mala1997heat}
\bibinfo{author}{G.~M. Mala}, \bibinfo{author}{D.~Li}, \bibinfo{author}{J.~D.
  Dale}, \bibinfo{title}{Heat transfer and fluid flow in microchannels},
  {\color{red} \bibinfo{journal}{International Journal of Heat and Mass
  Transfer}} \bibinfo{volume}{40}~(\bibinfo{number}{13})
  (\bibinfo{year}{1997}{\natexlab{b}}) \bibinfo{pages}{3079--3088}.

\bibitem[{Chun and Kwak(2003)}]{chun2003electrokinetic}
\bibinfo{author}{M.-S. Chun}, \bibinfo{author}{H.-W. Kwak},
  \bibinfo{title}{Electrokinetic flow and electroviscous effect in a charged
  slit-like microfluidic channel with nonlinear Poisson-Boltzmann field},
  {\color{red} \bibinfo{journal}{Korea-Australia Rheology Journal}}
  \bibinfo{volume}{15}~(\bibinfo{number}{2}) (\bibinfo{year}{2003})
  \bibinfo{pages}{83--90}.

\bibitem[{Ren and Li(2004)}]{ren2004electroviscous}
\bibinfo{author}{C.~L. Ren}, \bibinfo{author}{D.~Li},
  \bibinfo{title}{Electroviscous effects on pressure-driven flow of dilute
  electrolyte solutions in small microchannels}, {\color{red}
  \bibinfo{journal}{Journal of Colloid and Interface Science}}
  \bibinfo{volume}{274}~(\bibinfo{number}{1}) (\bibinfo{year}{2004})
  \bibinfo{pages}{319--330}.

\bibitem[{Chen et~al.(2004)Chen, Toh, Chai, and Yang}]{chen2004developing}
\bibinfo{author}{X.~Chen}, \bibinfo{author}{K.~Toh}, \bibinfo{author}{J.~Chai},
  \bibinfo{author}{C.~Yang}, \bibinfo{title}{Developing pressure-driven liquid
  flow in microchannels under the electrokinetic effect}, {\color{red}
  \bibinfo{journal}{International Journal of Engineering Science}}
  \bibinfo{volume}{42}~(\bibinfo{number}{5-6}) (\bibinfo{year}{2004})
  \bibinfo{pages}{609--622}.

\bibitem[{Joly et~al.(2006)Joly, Ybert, Trizac, and Bocquet}]{joly2006liquid}
\bibinfo{author}{L.~Joly}, \bibinfo{author}{C.~Ybert},
  \bibinfo{author}{E.~Trizac}, \bibinfo{author}{L.~Bocquet},
  \bibinfo{title}{Liquid friction on charged surfaces: From hydrodynamic
  slippage to electrokinetics}, {\color{red} \bibinfo{journal}{The Journal of
  Chemical Physics}} \bibinfo{volume}{125}~(\bibinfo{number}{20})
  (\bibinfo{year}{2006}) \bibinfo{pages}{204716}.

\bibitem[{Xuan(2008)}]{xuan2008streaming}
\bibinfo{author}{X.~Xuan}, \bibinfo{title}{Streaming potential and
  electroviscous effect in heterogeneous microchannels}, {\color{red}
  \bibinfo{journal}{Microfluidics and Nanofluidics}}
  \bibinfo{volume}{4}~(\bibinfo{number}{5}) (\bibinfo{year}{2008})
  \bibinfo{pages}{457--462}.

\bibitem[{Wang and Wu(2010)}]{wang2010flow}
\bibinfo{author}{L.~Wang}, \bibinfo{author}{J.~Wu}, \bibinfo{title}{Flow
  behavior in microchannel made of different materials with wall slip velocity
  and electro-viscous effects}, {\color{red} \bibinfo{journal}{Acta Mechanica
  Sinica}} \bibinfo{volume}{26}~(\bibinfo{number}{1}) (\bibinfo{year}{2010})
  \bibinfo{pages}{73--80}.

\bibitem[{Jamaati et~al.(2010)Jamaati, Niazmand, and
  Renksizbulut}]{jamaati2010pressure}
\bibinfo{author}{J.~Jamaati}, \bibinfo{author}{H.~Niazmand},
  \bibinfo{author}{M.~Renksizbulut}, \bibinfo{title}{Pressure-driven
  electrokinetic slip-flow in planar microchannels}, {\color{red}
  \bibinfo{journal}{International Journal of Thermal Sciences}}
  \bibinfo{volume}{49}~(\bibinfo{number}{7}) (\bibinfo{year}{2010})
  \bibinfo{pages}{1165--1174}.

\bibitem[{Zhao and Yang(2011)}]{zhao2011competition}
\bibinfo{author}{C.~Zhao}, \bibinfo{author}{C.~Yang}, \bibinfo{title}{On the
  competition between streaming potential effect and hydrodynamic slip effect
  in pressure-driven microchannel flows}, {\color{red}
  \bibinfo{journal}{Colloids and Surfaces A: Physicochemical and Engineering
  Aspects}} \bibinfo{volume}{386}~(\bibinfo{number}{1-3})
  (\bibinfo{year}{2011}) \bibinfo{pages}{191--194}.

\bibitem[{Tan and Liu(2014)}]{tan2014combined}
\bibinfo{author}{D.~Tan}, \bibinfo{author}{Y.~Liu}, \bibinfo{title}{Combined
  effects of streaming potential and wall slip on flow and heat transfer in
  microchannels}, {\color{red} \bibinfo{journal}{International Communications
  in Heat and Mass Transfer}} \bibinfo{volume}{53} (\bibinfo{year}{2014})
  \bibinfo{pages}{39--42}.

\bibitem[{Jing and Bhushan(2015)}]{jing2015electroviscous}
\bibinfo{author}{D.~Jing}, \bibinfo{author}{B.~Bhushan},
  \bibinfo{title}{Electroviscous effect on fluid drag in a microchannel with
  large zeta potential}, {\color{red} \bibinfo{journal}{Beilstein Journal of
  Nanotechnology}} \bibinfo{volume}{6}~(\bibinfo{number}{1})
  (\bibinfo{year}{2015}) \bibinfo{pages}{2207--2216}.

\bibitem[{Matin and Khan(2016)}]{matin2016electrokinetic}
\bibinfo{author}{M.~H. Matin}, \bibinfo{author}{W.~A. Khan},
  \bibinfo{title}{Electrokinetic effects on pressure driven flow of
  viscoelastic fluids in nanofluidic channels with Navier slip condition},
  {\color{red} \bibinfo{journal}{Journal of Molecular Liquids}}
  \bibinfo{volume}{215} (\bibinfo{year}{2016}) \bibinfo{pages}{472--480}.

\bibitem[{Jing et~al.(2017)Jing, Pan, and Wang}]{jing2017non}
\bibinfo{author}{D.~Jing}, \bibinfo{author}{Y.~Pan}, \bibinfo{author}{X.~Wang},
  \bibinfo{title}{The non-monotonic overlapping EDL-induced electroviscous
  effect with surface charge-dependent slip and its size dependence},
  {\color{red} \bibinfo{journal}{International Journal of Heat and Mass
  Transfer}} \bibinfo{volume}{113} (\bibinfo{year}{2017})
  \bibinfo{pages}{32--39}.

\bibitem[{Matin(2017)}]{matin2017electroviscous}
\bibinfo{author}{M.~H. Matin}, \bibinfo{title}{Electroviscous effects on
  thermal transport of electrolytes in pressure driven flow through nanoslit},
  {\color{red} \bibinfo{journal}{International Journal of Heat and Mass
  Transfer}} \bibinfo{volume}{106} (\bibinfo{year}{2017})
  \bibinfo{pages}{473--481}.

\bibitem[{Kim and Kim(2018)}]{kim2018analysis}
\bibinfo{author}{S.~I. Kim}, \bibinfo{author}{S.~J. Kim},
  \bibinfo{title}{Analysis of the electroviscous effects on pressure-driven
  flow in nanochannels using effective ionic concentrations}, {\color{red}
  \bibinfo{journal}{Microfluidics and Nanofluidics}}
  \bibinfo{volume}{22}~(\bibinfo{number}{1}) (\bibinfo{year}{2018})
  \bibinfo{pages}{12}.

\bibitem[{Sailaja et~al.(2019)Sailaja, Srinivas, and
  Sreedhar}]{sailaja2019electroviscous}
\bibinfo{author}{A.~Sailaja}, \bibinfo{author}{B.~Srinivas},
  \bibinfo{author}{I.~Sreedhar}, \bibinfo{title}{Electroviscous effect of power
  law fluids in a slit microchannel with asymmetric wall zeta potentials},
  {\color{red} \bibinfo{journal}{Journal of Mechanics}}
  \bibinfo{volume}{35}~(\bibinfo{number}{4}) (\bibinfo{year}{2019})
  \bibinfo{pages}{537--547}.

\bibitem[{Mo and Hu(2019)}]{mo2019electroviscous}
\bibinfo{author}{X.~Mo}, \bibinfo{author}{X.~Hu},
  \bibinfo{title}{Electroviscous effect on pressure driven flow and related
  heat transfer in microchannels with surface chemical reaction}, {\color{red}
  \bibinfo{journal}{International Journal of Heat and Mass Transfer}}
  \bibinfo{volume}{130} (\bibinfo{year}{2019}) \bibinfo{pages}{813--820}.

\bibitem[{Li et~al.(2021)Li, Liu, Liu, Feng, and Mo}]{li2021combined}
\bibinfo{author}{C.~Li}, \bibinfo{author}{Z.~Liu}, \bibinfo{author}{X.~Liu},
  \bibinfo{author}{Z.~Feng}, \bibinfo{author}{X.~Mo}, \bibinfo{title}{Combined
  effect of surface charge and boundary slip on pressure-driven flow and
  convective heat transfer in nanochannels with overlapping electric double
  layer}, {\color{red} \bibinfo{journal}{International Journal of Heat and Mass
  Transfer}} \bibinfo{volume}{176} (\bibinfo{year}{2021})
  \bibinfo{pages}{121353}.

\bibitem[{Li et~al.(2022)Li, Liu, Qiao, Feng, and Tian}]{li2022electroviscous}
\bibinfo{author}{C.~Li}, \bibinfo{author}{Z.~Liu}, \bibinfo{author}{N.~Qiao},
  \bibinfo{author}{Z.~Feng}, \bibinfo{author}{Z.~Q. Tian}, \bibinfo{title}{The
  electroviscous effect in nanochannels with overlapping electric double layers
  considering the height size effect on surface charge}, {\color{red}
  \bibinfo{journal}{Electrochimica Acta}} \bibinfo{volume}{419}
  (\bibinfo{year}{2022}) \bibinfo{pages}{140421}.

\bibitem[{Banerjee et~al.(2022)Banerjee, Pati, and
  Biswas}]{banerjee2022analysis}
\bibinfo{author}{D.~Banerjee}, \bibinfo{author}{S.~Pati},
  \bibinfo{author}{P.~Biswas}, \bibinfo{title}{Analysis of electroviscous
  effect and heat transfer for flow of non-Newtonian fluids in a microchannel
  with surface charge-dependent slip at high zeta potentials}, {\color{red}
  \bibinfo{journal}{Physics of Fluids}}
  \bibinfo{volume}{34}~(\bibinfo{number}{11}) (\bibinfo{year}{2022})
  \bibinfo{pages}{112016}.

\bibitem[{Yang et~al.(1998)Yang, Li, and Masliyah}]{yang1998modeling}
\bibinfo{author}{C.~Yang}, \bibinfo{author}{D.~Li}, \bibinfo{author}{J.~H.
  Masliyah}, \bibinfo{title}{Modeling forced liquid convection in rectangular
  microchannels with electrokinetic effects}, {\color{red}
  \bibinfo{journal}{International Journal of Heat and Mass Transfer}}
  \bibinfo{volume}{41}~(\bibinfo{number}{24}) (\bibinfo{year}{1998})
  \bibinfo{pages}{4229--4249}.

\bibitem[{Ren et~al.(2001)Ren, Li, and Qu}]{ren2001electro}
\bibinfo{author}{L.~Ren}, \bibinfo{author}{D.~Li}, \bibinfo{author}{W.~Qu},
  \bibinfo{title}{Electro-viscous effects on liquid flow in microchannels},
  {\color{red} \bibinfo{journal}{Journal of Colloid and Interface Science}}
  \bibinfo{volume}{233}~(\bibinfo{number}{1}) (\bibinfo{year}{2001})
  \bibinfo{pages}{12--22}.

\bibitem[{Hsu et~al.(2002)Hsu, Kao, Tseng, and Chen}]{hsu2002electrokinetic}
\bibinfo{author}{J.-P. Hsu}, \bibinfo{author}{C.-Y. Kao},
  \bibinfo{author}{S.~Tseng}, \bibinfo{author}{C.-J. Chen},
  \bibinfo{title}{Electrokinetic flow through an elliptical microchannel:
  Effects of aspect ratio and electrical boundary conditions}, {\color{red}
  \bibinfo{journal}{Journal of Colloid and Interface Science}}
  \bibinfo{volume}{248}~(\bibinfo{number}{1}) (\bibinfo{year}{2002})
  \bibinfo{pages}{176--184}.

\bibitem[{Bharti et~al.(2008)Bharti, Harvie, and Davidson}]{bharti2008steady}
\bibinfo{author}{R.~P. Bharti}, \bibinfo{author}{D.~J.~E. Harvie},
  \bibinfo{author}{M.~R. Davidson}, \bibinfo{title}{Steady flow of ionic liquid
  through a cylindrical microfluidic contraction--expansion pipe:
  Electroviscous effects and pressure drop}, {\color{red}
  \bibinfo{journal}{Chemical Engineering Science}}
  \bibinfo{volume}{63}~(\bibinfo{number}{14}) (\bibinfo{year}{2008})
  \bibinfo{pages}{3593--3604}.

\bibitem[{Davidson et~al.(2010)Davidson, Bharti, and
  Harvie}]{davidson2010electroviscous}
\bibinfo{author}{M.~R. Davidson}, \bibinfo{author}{R.~P. Bharti},
  \bibinfo{author}{D.~J.~E. Harvie}, \bibinfo{title}{Electroviscous effects in
  a Carreau liquid flowing through a cylindrical microfluidic contraction},
  {\color{red} \bibinfo{journal}{Chemical Engineering Science}}
  \bibinfo{volume}{65}~(\bibinfo{number}{23}) (\bibinfo{year}{2010})
  \bibinfo{pages}{6259--6269}.

\bibitem[{Berry et~al.(2011)Berry, Davidson, Bharti, and
  Harvie}]{berry2011effect}
\bibinfo{author}{J.~Berry}, \bibinfo{author}{M.~Davidson},
  \bibinfo{author}{R.~P. Bharti}, \bibinfo{author}{D.~Harvie},
  \bibinfo{title}{Effect of wall permittivity on electroviscous flow through a
  contraction}, {\color{red} \bibinfo{journal}{Biomicrofluidics}}
  \bibinfo{volume}{5}~(\bibinfo{number}{4}) (\bibinfo{year}{2011})
  \bibinfo{pages}{044102}.

\bibitem[{Davidson et~al.(2008)Davidson, Bharti, Liovic, and
  Harvie}]{davidson2008electroviscous}
\bibinfo{author}{M.~R. Davidson}, \bibinfo{author}{R.~P. Bharti},
  \bibinfo{author}{P.~Liovic}, \bibinfo{author}{D.~J. Harvie},
  \bibinfo{title}{Electroviscous effects in low Reynolds number flow through a
  microfluidic contraction with rectangular cross-section}, {\color{red} in:
  {\color{red} \bibinfo{booktitle}{Proceedings of World Academy of Science,
  Engineering and Technology}}}, vol.~\bibinfo{volume}{30},
  \bibinfo{pages}{256--260}, \bibinfo{year}{2008}.

\bibitem[{Dhakar and Bharti(2023{\natexlab{b}})}]{dhakar2023the}
\bibinfo{author}{J.~Dhakar}, \bibinfo{author}{R.~P. Bharti},
  \bibinfo{title}{{CFD} analysis of the influence of contraction size on
  electroviscous flow through the slit-type non-uniform microfluidic device},
  {\color{red} \bibinfo{journal}{To be communciated}} .

\bibitem[{Dhakar and Bharti(2022{\natexlab{b}})}]{dhakar2022slip}
\bibinfo{author}{J.~Dhakar}, \bibinfo{author}{R.~P. Bharti},
  \bibinfo{title}{Slip Effects in Ionic Liquids Flow Through a
  Contraction--Expansion Microfluidic Device}, in: \bibinfo{editor}{R.~P.
  Bharti}, \bibinfo{editor}{K.~Gangawane} (Eds.), {\color{red}
  \bibinfo{booktitle}{Recent Trends in Fluid Dynamics Research}},
  chap.~\bibinfo{chapter}{12}, \bibinfo{publisher}{Springer},
  \bibinfo{pages}{149--159}, \bibinfo{year}{2022}{\natexlab{b}}.

\bibitem[{Ghosal(2003)}]{ghosal2003effect}
\bibinfo{author}{S.~Ghosal}, \bibinfo{title}{The effect of wall interactions in
  capillary-zone electrophoresis}, {\color{red} \bibinfo{journal}{journal of
  Fluid Mechanics}} \bibinfo{volume}{491} (\bibinfo{year}{2003})
  \bibinfo{pages}{285--300}.

\bibitem[{Ajdari(1996)}]{ajdari1996generation}
\bibinfo{author}{A.~Ajdari}, \bibinfo{title}{Generation of transverse fluid
  currents and forces by an electric field: electro-osmosis on charge-modulated
  and undulated surfaces}, {\color{red} \bibinfo{journal}{Physical Review E}}
  \bibinfo{volume}{53}~(\bibinfo{number}{5}) (\bibinfo{year}{1996})
  \bibinfo{pages}{4996}.

\bibitem[{Jain and Nandakumar(2013)}]{jain2013optimal}
\bibinfo{author}{M.~Jain}, \bibinfo{author}{K.~Nandakumar},
  \bibinfo{title}{Optimal patterning of heterogeneous surface charge for
  improved electrokinetic micromixing}, {\color{red}
  \bibinfo{journal}{Computers \& Chemical Engineering}} \bibinfo{volume}{49}
  (\bibinfo{year}{2013}) \bibinfo{pages}{18--24}.

\bibitem[{Bhattacharyya and Bag(2019)}]{bhattacharyya2019enhanced}
\bibinfo{author}{S.~Bhattacharyya}, \bibinfo{author}{N.~Bag},
  \bibinfo{title}{Enhanced electroosmotic flow of Herschel-Bulkley fluid in a
  channel patterned with periodically arranged slipping surfaces}, {\color{red}
  \bibinfo{journal}{Physics of Fluids}}
  \bibinfo{volume}{31}~(\bibinfo{number}{7}) (\bibinfo{year}{2019})
  \bibinfo{pages}{072007}.

\bibitem[{Nayak et~al.(2018)Nayak, Banerjee, and Weigand}]{nayak2018mixing}
\bibinfo{author}{A.~K. Nayak}, \bibinfo{author}{A.~Banerjee},
  \bibinfo{author}{B.~Weigand}, \bibinfo{title}{Mixing and charge transfer in a
  nanofluidic system due to a patterned surface}, {\color{red}
  \bibinfo{journal}{Applied Mathematical Modelling}} \bibinfo{volume}{54}
  (\bibinfo{year}{2018}) \bibinfo{pages}{483--501}.

\bibitem[{Chu and Jian(2019)}]{chu2019magnetohydrodynamic}
\bibinfo{author}{X.~Chu}, \bibinfo{author}{Y.~Jian},
  \bibinfo{title}{Magnetohydrodynamic electro-osmotic flow of Maxwell fluids
  with patterned charged surface in narrow confinements}, {\color{red}
  \bibinfo{journal}{Journal of Physics D: Applied Physics}}
  \bibinfo{volume}{52}~(\bibinfo{number}{40}) (\bibinfo{year}{2019})
  \bibinfo{pages}{405003}.

\bibitem[{Guan et~al.(2021)Guan, Yang, and Wu}]{guan2021mixing}
\bibinfo{author}{Y.~Guan}, \bibinfo{author}{T.~Yang}, \bibinfo{author}{J.~Wu},
  \bibinfo{title}{Mixing and transport enhancement in microchannels by
  electrokinetic flows with charged surface heterogeneity}, {\color{red}
  \bibinfo{journal}{Physics of Fluids}}
  \bibinfo{volume}{33}~(\bibinfo{number}{4}) (\bibinfo{year}{2021})
  \bibinfo{pages}{042006}.

\bibitem[{Ghosal(2006)}]{ghosal2006electrokinetic}
\bibinfo{author}{S.~Ghosal}, \bibinfo{title}{Electrokinetic flow and dispersion
  in capillary electrophoresis}, {\color{red} \bibinfo{journal}{Annu. Rev.
  Fluid Mech.}} \bibinfo{volume}{38} (\bibinfo{year}{2006})
  \bibinfo{pages}{309--338}.

\bibitem[{Ng and Zhou(2012)}]{ng2012dispersion}
\bibinfo{author}{C.-O. Ng}, \bibinfo{author}{Q.~Zhou},
  \bibinfo{title}{Dispersion due to electroosmotic flow in a circular
  microchannel with slowly varying wall potential and hydrodynamic slippage},
  {\color{red} \bibinfo{journal}{Physics of Fluids}}
  \bibinfo{volume}{24}~(\bibinfo{number}{11}) (\bibinfo{year}{2012})
  \bibinfo{pages}{112002}.

\bibitem[{Azari et~al.(2020)Azari, Sadeghi, and
  Chakraborty}]{azari2020electroosmotic}
\bibinfo{author}{M.~Azari}, \bibinfo{author}{A.~Sadeghi},
  \bibinfo{author}{S.~Chakraborty}, \bibinfo{title}{Electroosmotic flow and
  heat transfer in a heterogeneous circular microchannel}, {\color{red}
  \bibinfo{journal}{Applied Mathematical Modelling}} \bibinfo{volume}{87}
  (\bibinfo{year}{2020}) \bibinfo{pages}{640--654}.

\bibitem[{Dhakar and Bharti(2023{\natexlab{c}})}]{dhakar2023analysis}
\bibinfo{author}{J.~Dhakar}, \bibinfo{author}{R.~P. Bharti},
  \bibinfo{title}{{CFD} analysis of electroviscous effects in electrolyte
  liquid flow through heterogeneously charged uniform microfluidic device},
  {\color{red} \bibinfo{journal}{To be communicated}} .

\bibitem[{Harvie et~al.(2012)Harvie, Biscombe, and
  Davidson}]{harvie2012microfluidic}
\bibinfo{author}{D.~J. Harvie}, \bibinfo{author}{C.~J. Biscombe},
  \bibinfo{author}{M.~R. Davidson}, \bibinfo{title}{Microfluidic circuit
  analysis I: Ion current relationships for thin slits and pipes}, {\color{red}
  \bibinfo{journal}{Journal of Colloid and Interface Science}}
  \bibinfo{volume}{365}~(\bibinfo{number}{1}) (\bibinfo{year}{2012})
  \bibinfo{pages}{1--15}.

\bibitem[{Davidson et~al.(2016)Davidson, Berry, Pillai, and
  Harvie}]{davidson2016numerical}
\bibinfo{author}{M.~R. Davidson}, \bibinfo{author}{J.~D. Berry},
  \bibinfo{author}{R.~Pillai}, \bibinfo{author}{D.~J. Harvie},
  \bibinfo{title}{Numerical simulation of two-fluid flow of electrolyte
  solution with charged deforming interfaces}, {\color{red}
  \bibinfo{journal}{Applied Mathematical Modelling}}
  \bibinfo{volume}{40}~(\bibinfo{number}{3}) (\bibinfo{year}{2016})
  \bibinfo{pages}{1989--2001}.

\bibitem[{Sisavath et~al.(2002)Sisavath, Jing, Pain, and
  Zimmerman}]{Sisavath2002}
\bibinfo{author}{S.~Sisavath}, \bibinfo{author}{X.~Jing},
  \bibinfo{author}{C.~C. Pain}, \bibinfo{author}{R.~W. Zimmerman},
  \bibinfo{title}{Creeping flow through an axisymmetric sudden contraction or
  expansion}, {\color{red} \bibinfo{journal}{Journal of Fluids Engineering}}
  \bibinfo{volume}{124}~(\bibinfo{number}{1}) (\bibinfo{year}{2002})
  \bibinfo{pages}{273 -- 278}.

\bibitem[{Pimenta et~al.(2020)Pimenta, Toda‑Peters, Shen, Alves, and
  Haward}]{Pimenta2020}
\bibinfo{author}{F.~Pimenta}, \bibinfo{author}{K.~Toda‑Peters},
  \bibinfo{author}{A.~Q. Shen}, \bibinfo{author}{M.~A. Alves},
  \bibinfo{author}{S.~J. Haward}, \bibinfo{title}{Viscous flow through
  microfabricated axisymmetric contraction/expansion geometries}, {\color{red}
  \bibinfo{journal}{Experiments in Fluids}} \bibinfo{volume}{61}
  (\bibinfo{year}{2020}) \bibinfo{pages}{204}.

\end{thebibliography}
%
%
%
%
%
%
%
%
%
%
%
%
\end{document}